Roadmap

# Roadmap on Electronic Structure Codes in the Exascale Era

Vikram Gavini[1,37*], Stefano Baroni[2,3*], Volker Blum[4*], David R. Bowler[5,6*], Alexander Buccheri[7], James R. Chelikowsky[8,9,10*], Sambit Das[1], William Dawson[11], Pietro Delugas[2], Mehmet Dogan[9], Claudia Draxl[7*], Giulia Galli[12,13*], Luigi Genovese[14*], Paolo Giannozzi[3,15], Matteo Giantomassi[16], Xavier Gonze[16*], Marco Govoni[13,12*], Andris Gulans[17], François Gygi[18*], John M. Herbert[19*], Sebastian Kokott[20], Thomas D. Kühne[21,22*], Kai-Hsin Liou[8], Tsuyoshi Miyazaki[6], Phani Motamarri[23*], Ayako Nakata[6], John E. Pask[24], Christian Plessl[22,25], Laura E. Ratcliff[26], Ryan M. Richard[27,28], Mariana Rossi[29,30], Robert Schade[22,25], Matthias Scheffler[29], Ole Schütt[31], Phanish Suryanarayana[32*], Marc Torrent[33,34], Lionel Truflandier[35], Theresa L. Windus[28,27], Qimen Xu[32], Victor W.-Z. Yu[13], Danny Perez[36,37*].

[1] Department of Mechanical Engineering, University of Michigan, Ann Arbor, MI, 48109, USA
[2] SISSA-ISAS, Trieste, 34136 Trieste, Italy
[3] IOM-CNR, Trieste, 34136 Trieste, Italy
[4] Duke University, Durham, NC, 27708, USA
[5] London Centre for Nanotechnology, London WC1H 0AH, United Kingdom
[6] WPI-MANA, National Institute for Materials Science, Ibaraki 305-0044 JAPAN
[7] Humboldt-Universität zu Berlin, 10117 Berlin, Germany
[8] McKetta Department of Chemical Engineering, University of Texas at Austin, Austin, TX, 78712, USA
[9] Center for Computational Materials, Oden Institute, University of Texas at Austin, Austin, TX, 78712, USA
[10] Department of Physics, University of Texas at Austin, Austin, Tx, 78712, USA
[11] RIKEN Center for Computational Science, Kobe, Hyogo, 650-0047, Japan
[12] Pritzker School of Molecular Engineering, University of Chicago, Chicago, IL, 60637, USA
[13] Materials Science Division, Argonne National Laboratory, Lemont, IL, 60439, USA
[14] Université Grenoble Alpes, MEM, L_Sim, F-38000 Grenoble, France
[15] University of Udine, 33100 Udine, Italy
[16] IMCN Université catholique de Louvain, Louvain-la-neuve, Belgium
[17] University of Latvia, Rīga, LV-1586, Latvia
[18] University of California, Davis, CA, 95616, USA
[19] Department of Chemistry & Biochemistry, The Ohio State University, Columbus, OH, 43210, USA
[20] Materials Simulations from First Principles e.V. (MS1P), 14195 Berlin, Germany
[21] Dynamics of Condensed Matter and Center for Sustainable Systems Design, Chair of Theoretical Chemistry, Paderborn University, D-33098 Germany
[22] Paderborn Center for Parallel Computing, Paderborn University, D-33098 Germany
[23] Department of Computational and Data Sciences, Indian Institute of Science, Bengaluru, Karnataka, 560012, India
[24] Lawrence Livermore National Laboratory, Livermore, CA, 94550, USA
[25] Department of Computer Science, Paderborn University, D-33098, Germany
[26] Centre for Computational Chemistry, School of Chemistry, University of Bristol, Bristol BS8 1TS, United Kingdom
[27] Ames National Laboratory, Ames, IA, 50011, USA
[28] Iowa State University, Ames, IA, 50011, USA
[29] The NOMAD Laboratory at the FHI of the Max-Planck-Gesellschaft and IRIS Adlershof of the Humboldt Universität zu Berlin, 14195 Berlin, Germany
[30] Max-Planck-Institute for the Structure and Dynamics of Matter, 22761 Hamburg, Germany
[31] CP2K Foundation, CH-8006 Zurich Switzerland

[32] College of Engineering, Georgia Institute of Technology, Atlanta, GA, 30332, USA
[33] CEA DAM-DIF, F-91297 Arpajon, France
[34] Université Paris-Saclay, CEA, Laboratoire Matière en Conditions Extrêmes, 91680 Bruyères-le-Châtel, France
[35] Institut des Sciences Moleculaires, Université Bordeaux, 33405 TALENCE cedex, France
[36] Theoretical Division T-1, Los Alamos National Laboratory, Los Alamos, NM, 87545, USA

[37] Lead author(s) of the Roadmap, to whom any correspondence should be addressed.

*Lead author of each section.

E-mails: vikramg@umich.edu; danny_perez@lanl.gov.

## Abstract

Electronic structure calculations have been instrumental in providing many important insights into a range of physical and chemical properties of various molecular and solid-state systems. Their importance to various fields, including materials science, chemical sciences, computational chemistry and device physics, is underscored by the large fraction of available public supercomputing resources devoted to these calculations. As we enter the exascale era, exciting new opportunities to increase simulation numbers, sizes, and accuracies present themselves. In order to realize these promises, the community of electronic structure software developers will however first have to tackle a number of challenges pertaining to the efficient use of new architectures that will rely heavily on massive parallelism and hardware accelerators. This roadmap provides a broad overview of the state-of-the-art in electronic structure calculations and of the various new directions being pursued by the community. It covers 14 electronic structure codes, presenting their current status, their development priorities over the next five years, and their plans towards tackling the challenges and leveraging the opportunities presented by the advent of exascale computing.

## Contents

## Introduction

Since the deployment of the Frontier supercomputer at Oak Ridge National Laboratory in the United States in May of 2022, the high-performance computing world has officially entered the exascale era. Frontier can indeed deliver in excess of $10^{18}$ double-precision floating point operations per second. This milestone marks the transition from the 14-year long petascale era that began with the 2008 deployment of the Roadrunner supercomputer in Los Alamos National Laboratory in the United States. The prodigious increase in available computer power has dramatically expanded the space of possible simulations in terms of the sizes of simulations that can be executed, of the accessible simulation times, and of the physical fidelity/complexity of the simulations.

This immense potential for scientific discovery is however contingent on learning to efficiently harness the power of exascale computers. While for many years if was possible to reap the benefits from the constant increase in the transistor frequency and density in conventional CPU architectures without significant changes in the simulation codes, it will become increasingly difficult to do so in the exascale era where heterogeneous architectures with many-core CPUs coexisting with accelerators such as GPUs will be ubiquitous. Indeed, as of June 2022, 7 of the 10 most powerful computer systems are heterogeneous systems where the majority of the computing power is provided by accelerators. The massive parallelism inherent to these systems (Frontier has almost 9 million hardware cores), as well as the heterogeneous nature of the computing hardware and of the memory hierarchies, poses significant challenges to application developers wishing to leverage exascale computing. While the number and geographic distribution of exascale systems may remain limited for some years, the technology driving these systems will quickly cascade down to smaller workhorse systems routinely used in academia and industry. Further, with more than 500 petaflop systems currently deployed worldwide, it can therefore be expected that hundreds of exascale systems are likely to become available in the next decade.

Given the difficulties in adapting computational methods and codes to these extreme-scale machines, we believe that the exascale milestone is an ideal opportunity to take stock of the materials modeling community's plans in developing powerful strategies to ensure that the exascale revolution fulfills its promises of producing high-value scientific insights that address pressing scientific questions. Given their central importance in contemporary computational materials science, electronic structure methods present an ideal case study to better understand the challenges and opportunities of the exascale. Indeed, electronic structure methods are some of the largest consumers of computing cycles worldwide. For example, at the National Energy Research Scientific Computing Center in the United States (which hosts the Top-10 system Perlmutter at the time of writing), more than 20% of the total available computing cycles are consumed by electronic structure calculations applied to a broad range of problems of importance to materials science, physics, biology, and chemistry. The methodological sophistication of electronic structure methods coupled with the complexity of the codes in which they are implemented (often containing hundreds of thousands of lines of code) also suggest that the community will be facing significant challenges in adapting to the new reality.

The goal of this roadmap is to survey the current status and plans of a number of electronic structure development efforts. Each team was asked to contribute their thoughts on three questions:

1. What is the current status of your code, and what niche does it occupy in the broader ecosystem?
2. What are your development priorities over the next 5 years?
3. How do you plan to address the challenges posed by new architectures and by the constant increase in parallelism? How will your code make use of exascale computing?

While by no means exhaustive, we believe that this exercise provides an important snapshot of the current state of mind of code and method developers upon which thousands of practitioners in industry and academia rely. We view this Roadmap as an opportunity for users of these codes to plan for possible extensions of their current research programs in the upcoming years, as well as for funding agencies and stakeholders to identify possible areas of opportunities that would make a large impact in the field. An important objective of this Roadmap is also to be broad and inclusive. The surveyed codes therefore currently operate at a range of computational scales, vary in the numerical methods they use, range from research codes that seek to showcase new avenues to extremely broadly used workhorse codes, etc. The objective is to gather a very broad set of possible visions of the future, which aim at different scientific and computational targets.

Through these contributions, we identified common themes in terms of development priorities and of possible strategies to leverage massively-parallel platforms. Seven broad themes emerge:
1. Adapt to heterogeneous architectures such as CPU+GPU nodes. As stated above, this is possibly the most pressing technological driver faced by code developers, as GPU architectures are not naturally suited to many of the basic building blocks of electronic structure codes, but the overwhelming majority of the computing power of large computers is likely to be provided by them. Adapting to GPUs requires developing new methods that can expose vast amounts of parallelism on each node, which can in turn demand a dramatic rethinking of the key algorithms.
2. Development of more efficient and scalable distributed algorithms. The exascale computers will be composed of a very large number of nodes (likely in the tens of thousands or more) that will not share a unified global memory. Developing algorithms that can independently operate on small subsections of memory will be critical to large-scale performance.
3. Development of advanced exchange-correlation functionals and of post-DFT methods. These methods, such as Rung 4 and 5 functionals, Random Phase Approximation, or machine-learned exchange-correlation functionals, are aimed towards improving the predictive accuracies of strongly-correlated materials systems that are at the center of many scientific and technologically important problems. Due to their high computational cost and algorithmic complexity, these methods are prime candidates to efficiently leverage massive computational resources.
4. Calculation of response functions. These quantities are critical to estimate a number of key quantities such as spectroscopic properties, susceptibilities, phonons, etc. The high computational complexity of these methods can expose additional opportunities for parallelization that would benefit from large computational resources.
5. Modularization and interoperability. Traditionally, electronic structure codes were mostly used in isolation. Increasingly, electronic structure codes are being integrated into more complex workflows that require the execution of vast numbers of small calculations, e.g. for high-throughput materials screening, to parameterize higher-scale models, or to train machine learning algorithms. Tightly integrating electronic structure codes into such workflows requires designing flexible Application Programming Interfaces (APIs) that facilitate code coupling, or the development of modularized capabilities that can be "mixed-and-matched" depending on the application use case. While no single electronic structure calculation necessarily runs at large scale, such workflows can expose massive opportunities for parallelization though large numbers of independent or weakly coupled calculations that can be executed simultaneously.
6. O(N) methods. The unfavorable scaling of traditional electronic structure methods strongly limits the system sizes that can be investigated. Indeed, a thousand-fold increase in computing power from the petascale to the exascale only results in a ten-fold increase in the number of electrons for cubically-scaling methods like density functional theory. Linear (or reduced) scaling approximations are key to truly allow for spatially-extended systems to be simulated directly.
7. Fast Molecular Dynamics. While classical molecular dynamics has proved to be extremely powerful at estimating a broad range of thermodynamic and dynamic properties, electronic structure calculations are typically limited to statics or to very short MD trajectories, which significantly limits their range of applicability. Developing methods to improve the simulation rate of ab initio MD would open the door to a number of opportunities to leverage large computers, including replica-based thermodynamic sampling algorithms for free energy calculations, or the investigation of quantum nuclear dynamics using path integral methods.

The following sections will also show that while common themes emerge, development teams have a different set of priorities that play on the core strength of their specific applications. This testifies to the vitality of dynamism of the field and promises exciting new opportunities for the users of these codes. While the electronic structure codes covered in the article are by no means exhaustive, we believe they are representative of the diverse ongoing efforts in the community towards the continuous development of mature codes as well as development of emerging codes.

**Acknowledgements**

DP acknowledges support from the Exascale Computing Project (17-SC-20-SC), a collaborative effort of the U.S. Department of Energy Office of Science and the National Nuclear Security Administration.

# 1 – Abinit

Xavier Gonze (1), Matteo Giantomassi (1), Marc Torrent (2,3)
1 IMCN Université catholique de Louvain, Louvain-la-neuve, Belgium
2 CEA DAM-DIF, F-91297 Arpajon, France
3 Université Paris-Saclay, CEA, Laboratoire Matière en Conditions Extrêmes, 91680 Bruyères-le-Châtel, France

**Background and Current Status**

When the ABINIT project started, 25 years ago, it was clear that the development of electronic structure calculations would have an enormous impact on condensed matter physics, while the variety of target properties (electronic, dynamical, dielectric, chemical, magnetic, etc.) required dozens of developers embarking for a long journey. ABINIT stemmed from the worldwide collaborative work of scientists embracing the free software philosophy. It is probably the first electronic structure software application released under an open-source license. Nowadays, as shown in Fig. 1, ABINIT relies on many different formalisms to address the properties of periodic solids, molecules and nanosystems: density functional theory (DFT), density-functional perturbation theory (DFPT), many-body perturbation theory (MBPT - GW and BSE), dynamical mean-field theory (DMFT), temperature-dependent effective potentials (TDEP) for anharmonic effects. Utilized by thousands of users worldwide, ABINIT takes part regularly in verification/validation efforts with other large electronic structure packages, and is interfaced with many mathematics/physics/IO libraries and data formats.

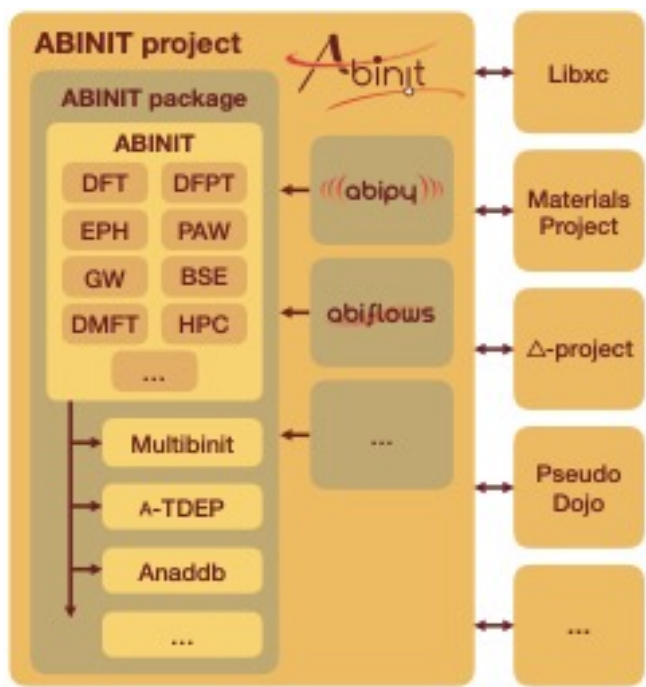

**Figure 1.** Components of the ABINIT project represented with respect to each other, or, on the right, to related projects. ([1], reprinted with permission)

ABINIT is particularly strong in DFPT, with several unique features (to our knowledge) such as flexoelectricity, dynamical quadrupole, strain perturbations, phonons under electric field, non-linear dielectric susceptibility and the electro-optic effect. Other capabilities of ABINIT might be available only in a few other codes, e.g. electron-phonon interaction (EPH), renormalization of gap, DMFT with forces, positron lifetime, DFPT for van der Waals solids, Raman cross-sections. ABINIT has been used for high-throughput computations (HTC) for several years already, and as a starting point for second-principles calculations (see recent generic publications [1,2]).

ABINIT also has a long track-record of parallelism for high-performance calculations (HPC). Before 2018, the code was essentially parallelized with distributed memory using MPI. Thanks to careful treatment of the Hamiltonian application, and relying on the LOBPCG algorithm [3], ABINIT could efficiently use several thousand processors. This algorithm has its own limits however. In order to achieve even better scalability, the "Chebyshev filtering" algorithm has been implemented, allowing use of more than 10000 processors [4].

**Development Priorities**

Multiple challenges lie ahead in terms of software development and methods, in particular because more complex levels of theory are needed in order to improve the predictive power of first-principles methods. The formalisms involved are much more memory and computationally intensive, hence exascale resources are needed to tackle these problems. The following priorities will be tackled by different groups within the ABINIT developer community.

Theories extending the DFT through the inclusion of the exact electronic exchange have appeared in the last two decades. One important priority is to go beyond ground-state (GS) properties by implementing response function calculations within DFPT with such exact electronic exchange-based functionals.

Another ambition is to apply accurate beyond-DFT methodologies based on Green's functions, --- specifically the GW approach --- to problems and systems currently out of reach with state-of-the-art computers and software. In a planewave-based code, the challenge is to reduce the cost of Fast Fourier Transforms (FFT), either by optimal use of the hardware (accelerators) or by employing new algorithms with lower complexity with respect to the system size.

An advanced ABINIT capability is to allow for predictive calculations of the properties of correlated electrons materials by coupling the DFT with the DMFT. In this case, the most expensive part is represented by the Quantum Monte-Carlo (QMC) algorithm whose scalability is therefore of crucial importance. One of our priorities is to limit its impact on the calculation while reducing the associated numerical noise.

Additionally, the increase in the system size as well as the reduction in time-to-solution can also be achieved by coupling ABINIT with more phenomenological approaches. Two methods are envisioned: second-principle methods employing effective potentials as implemented in the MULTIBINIT project [1], as well as coupling ABINIT with data-driven approaches based on efficient and transferable machine learning potentials. In both cases, the challenge is to maintain the accuracy of the underlying DFT method.

Finally, we would like to mention our developments in integrating ABINIT with the HTC paradigm. In the past years, ABINIT has been used for several HTC studies involving beyond GS properties such as quasi-particle energies within MBPT [5], optical properties [6], and also the phonons, dielectric constants and Born effective charges of 1521 semiconductors, made available on the materials-project website [7]. All these computations are significantly more expensive than standard GS applications and involve the execution of many complex workflows made of multiple interconnected tasks each of which has its own MPI parallelization scheme, computational load and memory requirements. For this reason, most of the high-level logic needed to drive these ABINIT workflows is implemented by a python driver that generates input files with optimized MPI input variables and orchestrates the submission of the different tasks while providing automatic error handlers and restart capabilities. A natural continuation of this effort is to extend the present implementation to automatically compute electron-phonon related quantities such as superconducting properties, temperature-dependent band structures, and transport properties including polaronic effects.

These new studies will require a high degree of synergy between the HTC infrastructure and the HPC part of ABINIT, as elucidated in [8] in which dynamical quadrupoles (third-order derivatives of the energy, recently implemented in the DFPT module) were found to be crucial for accurate computations of carrier mobilities in semiconductors.

**Meeting the Exascale Challenges**
Since 2018 the ABINIT roadmap to exascale is based on six pillars:

*(1) Improve the scalability of the diagonalization algorithm*
Any (standard) iterative diagonalization algorithm is based on two elementary building blocks: application of the Hamiltonian and a Rayleigh-Ritz procedure. Algorithms that exploit the first of these two operations more heavily should be preferred in the exascale context as they lead to less communication and memory access. For this reason, ABINIT now provides the Chebyshev filtering algorithm [4], and we plan to implement a "spectrum slicing" method [9] which would completely avoid the Rayleigh-Ritz step thus increasing parallel scalability.

*(2) Improved shared memory parallelism*
ABINIT was first designed with a *fine-grained* approach for shared memory parallelism in which low-level loops were parallelized with OpenMP directives. Now a *coarse-grained* model has been adopted in order to increase the work performed by each thread, minimize data movement, and also better connect physics to data mapping on different nodes. On recent HPC platforms, it is necessary to use GPU-based accelerators, implying commitment to a specific technology (proprietary software, directive-based languages, etc.).

*(3) Externalize the elementary kernels*
Most of the operations in a plane-wave code are based on elementary kernels such as linear algebra, matrix algebra or FFTs. All these operations are available either in mathematical libraries implemented by system vendors or in open-source projects that provide optimized and up-to-date implementations. The ELPA project of the MPCDF at the Fritz-Haber-Institute, used by ABINIT, is a noteworthy example. More extensive use of libraries will require refactoring some sections of code to leverage the different interfaces.

*(4) Low-level abstraction layer*
A significant portion of new high-level developments in ABINIT is usually done by researchers with a background in physics or chemistry rather than in computer science. To bridge the gap between the two worlds, we have introduced an abstraction layer exposing an easy-to-use

API to perform the typical tasks occurring in *ab initio* calculations while hiding the internal implementation. In the hidden layer, algorithms can be implemented in different languages and this allows computer scientists to fully exploit the capabilities of modern hardware, in particular graphics accelerators.

*(5) Ensure numerical stability*
On modern many-core architectures, codes like ABINIT may show numerical instabilities due to desynchronization effects. In other words, different cores operating on the same input values may produce slightly different results and this inconsistency may propagate through the parallel algorithm without (expensive) explicit synchronization operations. Vectorization specifically leads to unpredictable processes desynchronization yielding slightly different numerical results between runs. We use the *veritracer* tool [10] to identify critical sections that are very sensitive to numerical precision. A refactoring of these sections is planned.

*(6) Implement task management on heterogeneous architectures*
To run the code on heterogeneous architectures, like for instance large many-core nodes coupled to GPUs, we need different versions of the kernels for each kind of processing unit. With optimized task management, it will be possible to distribute the workload on each computing unit that will use its specific version. Achieving a good load balancing is one of the main technical challenges.

These six pillars should be considered for GS calculations with standard functionals as well as for more advanced functionals or for the other formalisms previously mentioned, thus improving all aspects of the parallel scalability of ABINIT. It should be noted, however, that ab initio studies are also becoming more and more complex and multiple calculations are usually needed to obtain the physical properties of interest. Fortunately, not all the steps of a typical ab initio workflow depend on each other and some of these jobs can be executed concurrently via e.g. a python manager. According to this philosophy, the focus is more on the embarrassingly parallel aspects by splitting ab initio computations into smaller tasks and then optimizing each task individually rather than executing the entire computation on the largest number of cores accessible. In our opinion, this represents a valuable complementary approach for taking advantage of exascale architectures.

**Concluding Remarks**
Over the years, ABINIT has evolved to include different formalisms enabling far-reaching applications: DFT, DFPT, many-body perturbation theory, DMFT, TDEP, etc. Nowadays, ABINIT can routinely use more than 10,000 processors in high-performance platforms for common DFT ground state calculations, and can be coupled with a high-throughput management of the numerous different tasks required by scientific workflows.

There is still room for improvements and expansion, in order to use a larger number of processors in the high-performance approach, both for common ground state calculations and for more advanced formalisms.

ABINIT roadmap to exascale is based on six pillars: (1) Improve the scalability of the diagonalization algorithm (2) Efficiently use the shared memory (3) Externalize the elementary operations as kernels (4) Add a low-level abstraction layer (5) Ensure numerical stability (6) Implement task management on heterogeneous architectures.

Reaching the exascale will also need complementary improvements in high-throughput workflows.

In addition to these axes, ABINIT build system and documentation will be improved, especially for exascale architectures. Indeed, the production of an optimized ABINIT executable needs to link advanced libraries (e.g. ELPA, SCALAPACK, libraries for GPU) and use specialized compilation options (e.g. OpenMP in conjunction with threaded libraries for BLAS/FFT). Few users are expert in such technicalities, and a suboptimal version of ABINIT might be built. The integration with package managers such as EasyBuild and Spack will be improved, so that system administrators can easily build and deploy different optimized versions (pure-MPI, MPI+OpenMP, GPU, etc.).

**Acknowledgements**
We thank the numerous contributors to the ABINIT project over the years, many of which are the authors of Refs. [1,2]. We also thank Matthieu Verstraete, Gian-Marco Rignanese, and Guillaume Colin de Verdière for careful reading of the manuscript, and associated suggestions. The development of the HPC aspects of ABINIT was made possible thanks to access to the resources of the french *Très Grand Centre de Calcul du CEA*. We gratefully thank the computing center teams for their availability and expert advice. The work has received partial support from the European Union's Horizon 2020 research and innovation programme, grant agreement N° 951786 through the Center of Excellence NOMAD (NOMAD CoE).

## 2 – BigDFT: Exploiting Novel Wavelet-Based Approaches for Sophisticated Computational Workflows Involving Large Systems

Laura E. Ratcliff, Centre for Computational Chemistry, School of Chemistry, University of Bristol, Bristol BS8 1TS, United Kingdom

William Dawson, RIKEN Center for Computational Science, Kobe, Hyogo, 650-0047, Japan

Luigi Genovese (luigi.genovese@cea.fr), Univ. Grenoble Alpes, MEM, L_Sim, F-38000 Grenoble, France

### 1 Background and Current Status

Starting in 2005, the BigDFT project aimed to test the advantages of Daubechies wavelets as a basis set for Kohn-Sham density functional theory (KS-DFT) using pseudopotentials. This led to the creation of the BigDFT code, which has optimal features of flexibility, performance and precision [1]. Furthermore, the wavelet-based approach has enabled the implementation of an algorithm for DFT calculations of large systems containing many thousands of atoms, with a computational effort which scales linearly with the number of atoms. In this contribution we show how the localised description of the KS problem, emerging from the features of the basis set, can provide a simplified description of large scale electronic structure calculations. This in turn enables the extraction of first-principles derived quantities which can characterise the electronic structure of systems which were impractical to simulate even very recently.

In BigDFT, the KS orbitals may be expressed either directly in wavelets (cubic scaling), or as a linear combination of *intermediate* basis functions (linear scaling), also referred to as support functions (SFs), where the SFs are strictly localised numerical functions represented in the wavelet basis. This intermediate basis set approach has been developed in a number of different codes, though BigDFT's use of wavelets as the underlying basis is unique.

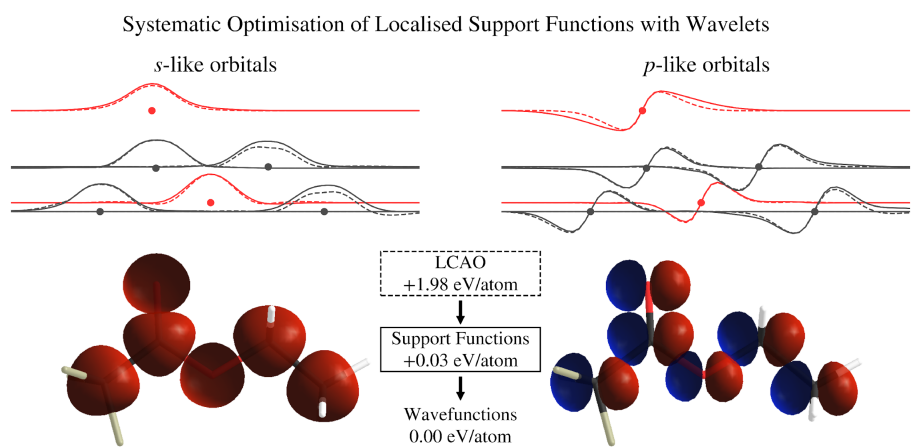

Figure 1: 3D isosurfaces of select *s*- and *p*-like optimised support functions for C (grey) and O (red) atoms in ethyl trifluoroacetate, with H and F atoms depicted in white and beige respectively. Also shown is a cross sectional representation with circles indicating the atomic positions, comparing the *in-situ* optimised support functions (solid lines) with the linear combination of atomic orbitals (LCAO) basis (dashed lines) which is used as an input guess. The changes, apparently subtle to-the-eye, nonetheless lead to a significant reduction in error relative to the full cubic scaling reference energy, where no localisation constraints are imposed on the KS wavefunctions.

The SF basis of BigDFT facilitates a linear scaling approach, while also offering numerous benefits for linear algebra based bottlenecks. While the SFs are initialised to atomic orbitals, their numerical representation allows for *in situ* optimisation, giving the accuracy of a large basis with only a minimal number of basis functions (see Fig. 1), and thus having small matrices even for large systems. The use of strictly localized, quasi-orthogonal basis functions further ensures that the matrices are sparse and well conditioned; the overlap matrix has a low spectral width and the ratio of the band gap to the spectral width of the Hamiltonian is relatively high. These properties lead to huge efficiency gains when using diagonalization-free methods as implemented in our CheSS [2] and NTPoly libraries [3].

BigDFT is now a mature and reliable code with a wide variety of features. A comprehensive overview of its capabilities is available in our recent review [1]. BigDFT uses dual space Gaussian type norm-conserving pseudopotentials including those with non-linear core corrections. It is parallelized using a combination of MPI and OpenMP and has support for GPU acceleration. BigDFT's flexible Poisson solver can handle a number of different boundary conditions including free, wire, surface, and periodic (orthorhombic only). There is also support for implicit solvent and external electric fields. In the cubic scaling approach, BigDFT can compute hybrid functionals and time-dependent (TD) DFT. BigDFT can be routinely applied to large systems. For example, the calculation of a 12,000 atom protein system requires about 1.2 hours of wall-time on 16 nodes of the Irene-ROME supercomputer. This calculation can be further accelerated for systems composed of repeated sub-units using a fragment [4] approach for molecules, or in the case of extended systems, a pseudo-fragment approach [5]. BigDFT is free and open source software, made available under the GPL license.

## 2 Development Priorities

Although BigDFT is able to treat systems composed of thousands of atoms, these calculations remain computationally demanding. It is therefore unrealistic (if not unnecessary) to expect DFT calculations to replace commonly used force field methods, as a full statistical sampling of a system's configuration space remains expensive. It is thus crucial to develop: 1) analysis techniques which use the results of large scale DFT calculations to gain new kinds of insights into emergent properties; 2) methodologies for probing excited state properties which are not accessible with forcefields; and 3) tools for creating complex, multi-scale workflows [6].

### 2.1 Complexity Reduction

The need for analysis techniques for large systems has led to the development of our *complexity reduction framework* [7] which takes the converged density matrix and Hamiltonian, and uses them to decompose systems into coarse-grained fragments. This procedure is based on two metrics:
- the purity indicator which measures the quality of a fragment;
- the fragment bond order which quantifies inter-fragment interactions.

This framework can further be combined with an energy decomposition analysis to quantify the strength of chemical interactions between different fragments. As a whole, these metrics can be used to automatically partition a system into fragments, design embedding environments for QM/MM type approaches, and produce graph-like views of system interactions.

The complexity reduction framework has been exploited for applications in biochemistry, where a challenging task is to propose reliable and systematic strategies for modulating the stability of protein/ligand and protein/protein assemblies. Because of the size of these assemblies, researchers must either construct small model systems, or rely on empirical cost functions or forcefields to quantify interactions. BigDFT's ability to treat large systems has been successfully applied in the context of drug-design of peptidic inhibitors of the main protease of SARS-CoV-2[8]. In this study, MD snapshots were post-processed using BigDFT calculations on systems made up of over 7000 atoms. We have also proposed a sequential multi-scale Molecular Modeling/Quantum Mechanical, mMM/QM [9], simulation approach, wherein BigDFT was used to support the accuracy of MD simulations involving over 20,000 atoms, as well as give new insights. For both these studies, we developed representations of the simulation results as interaction graphs showing the details and magnitudes of the interactions (at the residue scale) responsible for the stability of the studied complexes (see Fig. 2).

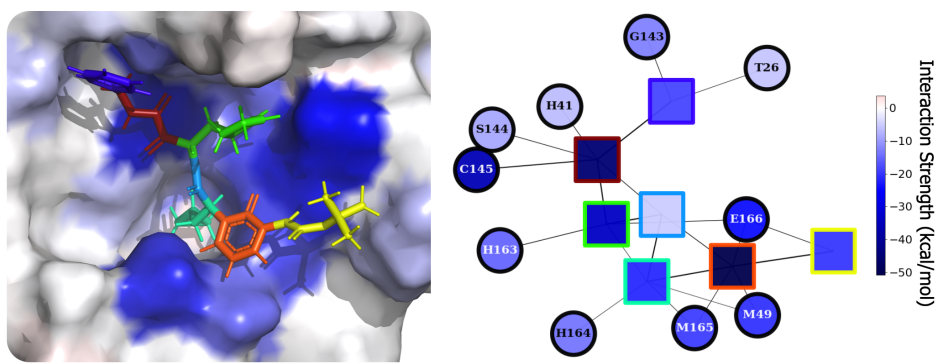

**Figure 2:** Interaction graph of a ketamide inhibitor (squares: molecular fragments) with the main protease of SARS-CoV-2 (circles: amino acids). The complexity reduction framework is used to measure the interaction strength (kcal/mol) of each inhibitor fragment with the amino acids of the protease.

## 2.2 Excited States

Another priority area for BigDFT is the development of excited state approaches, with a focus on methods which can be naturally combined with the fragment-based framework. To this end, constrained DFT (CDFT) has been implemented in such a way as to associate a charge constraint with a given fragment [4], enabling for example the calculation of charge transport parameters in a host-guest system typical of organic light emitting diodes (OLEDs) [10]. More recently, a new variant of CDFT has been developed, wherein an off-diagonal transition based constraint (T-CDFT) is applied, instead of a standard spatial constraint [11]. This allows the treatment of both local and charge-transfer (CT) like excitations, at a much lower computational cost than TDDFT, while also avoiding the severe underestimation of CT states seen with TDDFT and semi-local functionals. This approach is again compatible with the fragment-based framework, and could therefore be used in future works to take into account explicit environmental effects in excited state calculations of large and potentially disordered systems, such as OLED morphologies.

## 2.3 Workflow Management

When performing calculations of large systems, it is often necessary to apply more complicated workflows than are required at the small scale. Large systems take many steps to define and set up, multiple levels of theory might need to be coupled, calculations are deployed on remote supercomputers, and post-processing procedures are potentially costly in terms of processing power, memory, and disk space. To address these issues, we have developed a python package called PyBigDFT as a framework for managing DFT workflows. PyBigDFT is able to handle building complex systems or reading them from a variety of file types, performing calculations with BigDFT, and analysing calculation results (including through the complexity reduction framework). PyBigDFT further offers interoperability with other chemistry codes for multi-scale modelling. Using PyBigDFT, it is possible to create entire workflows inside a Jupyter notebook. This leads to not only reproducible calculations, but also computational *continuity*, where new research can be built on top of a existing results.

## 3  Meeting the Exascale Challenge

Looking ahead to the exascale era, BigDFT developments have focused on both internal code aspects (compilation and modularity, parallel performance), as well as external usability, both of which are necessary for an efficient transition to ever-evolving architectures. The compilation of the code suite relies on the splitting of the code components into modules, which are compiled by the bundler package. This package is defined from a fork of the Jhbuild program that has been conceived in the context of the GNOME developers consortium. This package lays the groundwork for developing a common infrastructure for compiling and linking together libraries for electronic structure codes, and it is employed as the basis for the ESL bundle. It can be used to install the BigDFT suite, as well as a variety of optional packages such as python modules that can be used with PyBigDFT.

Exascale machines will offer both the ability to simulate large systems as well as extraordinary capacity for high-throughput calculations. To facilitate driving thousands of calculations of large systems on those machines, we have been developing a *Remote Runner* capability in our PyBigDFT package. The remote runner automatically serializes data and arbitrary python closures, which are then executed through a supercomputer's queuing system. A remote function might be used to perform a BigDFT calculation, run other chemistry codes, or to perform resource heavy pre/post-processing steps. Calculations are performed in a lazy way, so that the first time a workflow is run the computationally demanding calculations are asynchronously submitted remotely, and subsequent runs of the workflow skip the calculation steps. This makes it easy to build analysis routines on top of the data generated from large scale DFT calculations.

Currently, BigDFT is being deployed on pre-exascale machines including Archer2 (Edinburgh Parallel Computing Centre), Fugaku (RIKEN Center for Computational Science), and Irene-ROME (Très Grand Centre de Calcul). We are hopeful that the ongoing developments of the BigDFT code will provide a roadmap for performing large scale quantum mechanical calculations of unprecedented size.

## 4  Acknowledgements

LER acknowledges support from an EPSRC Early Career Research Fellowship (EP/P033253/1). LG and WD acknowledge the joint CEA-RIKEN collaborative action. This work used computational resources of the supercomputer Fugaku provided by RIKEN through the HPCI System Research Project (Project ID: hp200179). LG acknowledges support from the MaX EU center of Excellence, and from French National computing resources (projects spe0011 and gen12049). We would like to thank Damien Caliste, Giuseppe Fisicaro, Louis Beal, Martina Stella, and Takahito Nakajima for discussions related to this manuscript, and everyone who has contributed to the BigDFT code over the years.

## 3 – The CONQUEST code: large scale and linear scaling DFT

D. R. Bowler[1,2], T. Miyazaki[2], A. Nakata[2] and L. Truflandier[3]
[1]London Centre for Nanotechnology, London, UK
[2]WPI-MANA, National Institute for Materials Science, Tsukuba, Japan
[3]Institut des Sciences Moleculaires, Université Bordeaux, Bordeaux, France

**Background and Current Status**

CONQUEST[1] is a DFT code which was designed from the beginning to enable extremely large-scale calculations on massively parallel platforms, implementing both exact and linear scaling solvers for the ground state.  It uses local basis sets (both pseudo-atomic orbitals, PAOs,[2] and systematically convergent B-splines[3]) and sparse matrix storage and operations to ensure locality in all aspects of the calculation.

Using exact diagonalisation approaches and a full PAO basis set, systems of up to 1,000 atoms can be modelled with relatively modest resources (200-500 cores), while use of multi-site support functions (MSSF)[4] enable calculations of up to 10,000 atoms with similar resources.  With linear scaling, the code demonstrates essentially perfect weak scaling (fixed atoms per process), and has been applied to over 1,000,000 atoms, scaling to nearly 200,000 cores[5]; it has been run on both the K computer and Fugaku, among other computers.

CONQUEST calculates the total energy, forces and stresses exactly, and allows structural optimisation of both ions and simulation cell.  Molecular dynamics calculations within the NVE, NVT and NPT ensembles are possible with both exact diagonalisation and linear scaling[6]. The code interfaces with LibXC to implement LDA and GGA functionals, with metaGGA and hybrid functionals under development.  Dispersion interactions can be included using semi-empirical methods (DFT-D2/3, TS) and vdW-DF.  The polarisation can be calculated using Resta's approach.

We have recently applied CONQUEST to calculations in complex ferroelectric systems with up to 5,000 atoms[7,8,9], investigating problems that require large simulation cells and electronic structure methods.  In Fig. 1a) and b), the local polarisation textures of PbTiO3 thin films on SrTiO3[7] are shown for two thicknesses of film: 9 layers (top) and 3 layers (bottom). The formation of polar vortices is clear in the thick film, while the thinner film cannot support these, instead showing a polar wave with chiral bubbles forming at the surface; we have extended these studies to investigate the interaction of domain walls with surface trenches[8]. In Fig 1c), we plot the partial charge density from the conduction band minimum (CBM) in YGaO3, superimposed on a map of the tilt angle of the GaO5 bipyramids relative to the (001) direction, which shows domain walls.  Domain wall meeting points (dark blue) are topologically protected, and show a concentration of the CBM, reflecting a reduced band gap[9].

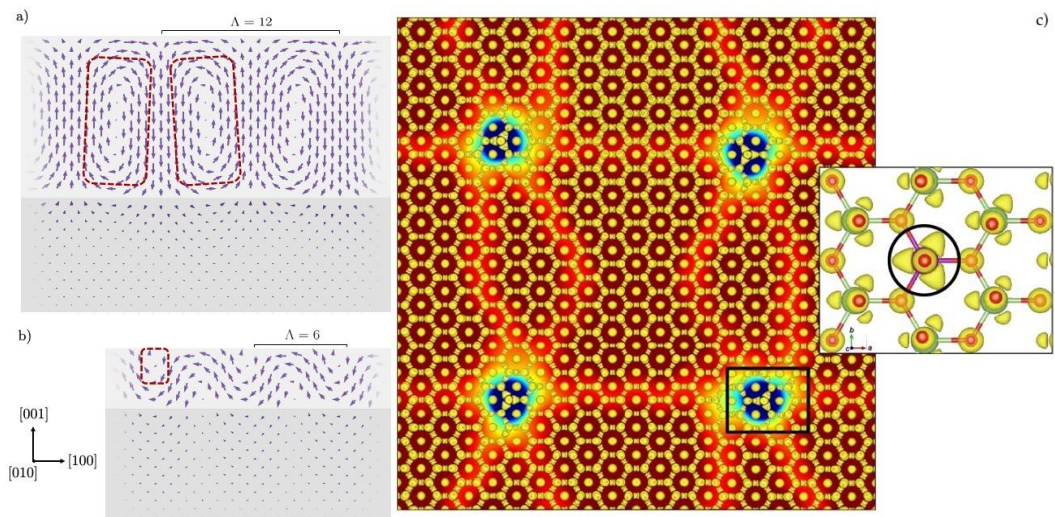

**Figure 1.** (a, b) Polarisation textures in PbTiO3 films on SrTiO3[7]. (c) Charge density from conduction band minimum in YGaO3, above a map of tilt angle in GaO5 bipyramids. Reproduced from Phys. Rev. B 102, 144103 (2020) with permission.

**Development Priorities**

CONQUEST already allows simulation of significantly larger systems than standard DFT codes, and we want to improve the efficiency and scaling of the code to enable even larger systems to be modelled. We also want to reduce the total computational time required per MD step, to enable longer timescales to be addressed. At present, exact diagonalisation simulations use the ScaLAPACK solver, but an interface to ELPA has been developed and is being deployed. Alongside these efficient approaches to exact diagonalisation, we are prioritising lower scaling solvers which allow the calculation of selected eigenstates.

Perhaps the best known of these in the electronic structure community is PEXSI, which is available through ELSI, though it would require some work to interface this to the CONQUEST matrix storage (which was designed specifically for high efficiency, highly parallel linear scaling calculations). We already have a post-hoc interface to an implementation of the Sakurai-Sugiura method, which scales as $O(mN^{1-2})$ when finding m eigenstates for N atoms, and shows extremely good parallel scaling, and we will investigate the possibility of incorporating this approach into CONQUEST as an alternative solver.

For the linear scaling solver, we will improve both the accuracy and the robustness. The key limitation on accuracy at present is the suitability of basis sets: the blip functions are accurate, and systematically improvable, but can be slow to converge, while PAOs are limited to relatively small sizes (typically single zeta plus polarisation, or SZP). The improvement of blip optimisation will concentrate on two aspects: first, the search methods used for the blips themselves; second, the integration with the linear scaling optimisation of the density matrix. For PAO basis sets, we will continue to develop an on-site equivalent of the MSSF approach, and develop an extension to full MSSF. We will also investigate alternative methods for linear scaling inversion of the overlap matrix, which is key to efficient linear scaling solution, and is sensitive to the basis set.

Efficient methods for metallic or small gap systems are also extremely important. The linear scaling solver in CONQUEST is not suitable for these systems, so we will investigate alternative approaches, including iterative methods. The question of large-scale, efficient solvers for these systems is one that is still of paramount importance to the large-scale DFT community as a whole.

We are in the process of including metaGGA functionals into CONQUEST, and will continue this alongside a robust, efficient implementation of exact exchange (for which we already have preliminary results). We will also enable solutions with spin-orbit coupling and the full Dirac equation. We have reported the successful linear scaling implementation of real-time TDDFT, and will extend this to the exact solvers, including the Casida linear-response approach for the exact solvers. We will also implement density-matrix perturbation theory (DMPT), as required by DFPT, for both linear scaling and exact solvers, to enable response function calculations to be performed.

**Meeting the Exascale Challenges**

The key challenge associated with the transition to exascale computing for CONQUEST is to update and adapt the code to new hardware architectures, in particular, efficient use of CPUs with many cores, and GPUs and other co-processors, while maintaining excellent parallel scaling. The code will need to become more heterogeneous, with on-node calculations distributed between CPU cores and GPUs, and a few MPI processes dedicated to inter-node communication; careful interleaving of different parts of the calculation to different hardware on a node will be key to maximising performance.

When working in exact diagonalisation mode, CONQUEST currently relies on ScaLAPACK, with an interface to ELPA in deployment. We will monitor developments in the area of exascale solvers (including projects such as MAGMA), as well as implementing other solvers which scale well in parallel, and which can make efficient use of local resources.

CONQUEST has, so far, shown no issues with parallel scaling when used in the weak scaling mode, even as far as 200,000 processes, as illustrated in Fig. 2. We are aware of areas in the code which might start to pose problems at larger process counts; the key area is the use of FFTs, though these are only used for the Hartree potential, calculation of gradients of the charge density, and for vdW-DF functionals. There are well-established alternatives for the first two of these operations which we will implement as necessary. The other area which may offer issues is the storage of atomic positions, velocities and forces, which are currently held globally on all processes, but can be made local (to the process responsible) if needed.

The overlap of communication and calculation is an approach that is inherently possible in much of the CONQUEST code[5], though has not been extensively implemented. This overlap fits well with multi-threading and heterogeneous hardware, and will be important to good on-node exascale performance. We will expand the areas of the code that are multithreaded (using OpenMP) as well as developing GPU implementations for numerically intensive parts of the calculation. At present, we have a preliminary GPU implementation for matrix multiplication, and a full GPU implementation of the O(N) solver would offer significant advantages. We will also test the off-loading of other parts, including force calculations and mapping from density matrix to charge density on the grid, along with the EXX implementation.

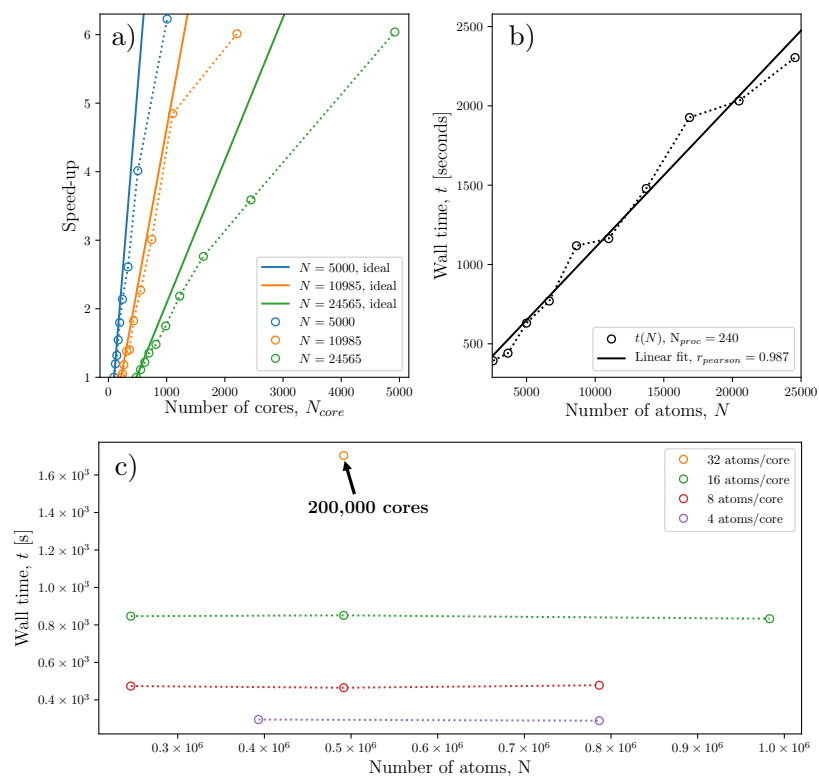

**Figure 2.** (a) Strong scaling for bulk PbTiO3 on the UK ARCHER computer; (b) linear scaling of computational time for the same system; (c) weak scaling on the K computer for systems up to 1,000,000 atoms of silicon. Reproduced from J. Chem. Phys. 152 164112 (2020) with permission.

**Concluding Remarks**

The search for new functional materials with complex structures is a key target for exa-scale computing, and the efficient use of massively parallel DFT calculations is an important part of this search. The CONQUEST code enables large-scale exact DFT simulations with relatively modest hardware resources, paving the way for large numbers of calculations on large simulations cells on exa-scale hardware. At the same time, with linear scaling, it is capable of modelling systems with many millions of atoms, hence applying DFT to systems of experimentally relevant size in many different disciplines. With improvements to solution time, long timescales and efficient structural relaxation will become widely available for these very large systems.

**Acknowledgements**

*The authors would like to acknowledge the many contributions of Professor Mike Gillan, who founded and led the CONQUEST project; we are also grateful to the other CONQUEST developers and users for all their input. The study relied on computational support from the UK Materials and Molecular Modelling Hub, which is partially funded by EPSRC (EP/P020194), for which access was obtained via the UKCP consortium and funded by EPSRC Grant Ref. No. EP/P022561/1. This work also used the ARCHER UK National Supercomputing Service funded by the UKCP consortium EPSRC Grant Ref. No. EP/P022561/1. This work is also partially supported by the projects JSPS Grant-in-Aid for Transformative Research Areas (A) "Hyper-Ordered Structures Science" (Grant Nos. 20H05883 and 20H05878) and JSPS Grant-in-Aid for Scientific Research (Grant Nos. 18H01143, 21H01008). Calculations were performed also on the Numerical Materials Simulator at NIMS in Japan.*

# 4 - CP2K on the road to exascale

**Thomas D. Kühne**[†,§,*], **Christian Plessl**[‡,§], **Robert Schade**[‡,§], **and Ole Schütt**[¶]

[†] Dynamics of Condensed Matter and Center for Sustainable Systems Design, Chair of Theoretical Chemistry, Paderborn University, D-33098 Germany
[‡] Department of Computer Science, Paderborn University, D-33098 Germany
[§] Paderborn Center for Parallel Computing, Paderborn University, D-33098 Germany
[¶] CP2K Foundation, CH-8006 Zurich Switzerland

E-mail: `tkuehne@cp2k.org`

**Abstract.**

The CP2K program package, which can be considered as the swiss army knife of atomistic simulations, is presented with a special emphasis on *ab-initio* molecular dynamics using the second-generation Car-Parrinello method. After outlining current and near-term development efforts with regards to massively parallel low-scaling post-Hartree-Fock and eigenvalue solvers, novel approaches on how we plan to take full advantage of future low-precision hardware architectures are introduced. Our focus here is on combining our submatrix method with the approximate computing paradigm to address the immanent exascale era.

## 1. Background and Current Status

The open-source simulation package CP2K is an extensive quantum chemistry and condensed matter physics program that comprises a large variety of different theoretical methods and computational approaches to conduct most diverse atomistic simulations for large-scale condensed-phase systems, such as liquids, solids, nanomaterials and molecular structures, to name just a few. All of this is made possible from the outset by the design of highly efficient algorithms with a focus on excellent parallel scalability and suitability for novel high-performance computing architectures, as demonstrated by its QUICKSTEP electronic structure module. The latter is based on the Gaussian and plane wave (GPW) approach and its all-electron variant, the Gaussian-augmented plane wave (GAPW) method, which allows for a particularly efficient treatment of large-scale orbital-free and Kohn–Sham density functional theory (DFT) calculations, as well as wavefunction-based correlation methods, such as Hartree-Fock (HF), second-order Møller–Plesset perturbation theory (MP2), random phase approximation (RPA), hybrid and double-hybrid DFT and the GW approximation, all with arbitrary boundary conditions [1].

However, the great appeal of CP2K lies in the possibility to combine all available total energy and force methods with a wide selection of sampling techniques, such as Monte Carlo, Ehrenfest/real-time dynamics and most importantly molecular dynamics, as well as advanced free-energy and rare-event sampling schemes, to enable realistic simulations at finite-temperature beside conventional static calculations. On the one hand, this necessitates the general availability of analytic gradients in particular for periodic boundary conditions, in order to permit the efficient calculation of nuclear forces. On the other hand, minimum time to solution is essential to allow for an extensive sampling via the techniques listed above. In that respect, a unique selling point of CP2K is the implementation of the second-generation Car-Parrinello method, which allows to routinely conduct nanosecond long DFT-based *ab-initio* molecular dynamics (AIMD) simulations with thousands of atoms. The superior efficiency of this approach originates from the design of an improved coupled electron-ion dynamics that keeps the electrons very close to their instantaneous ground-state using just one preconditioned gradient calculation per AIMD step, which can thus be thought of as an electronic force to propagate the electronic degrees of freedom in dimensionless time [2].

## 2. Development Priorities

Beside the implementation of sophisticated spectroscopic properties [3, 4], which are either based on density functional perturbation theory or time-dependent DFT within the Tamm-Dancoff approximation [1, 5], the current development priorities are focused mainly on devising novel low-scaling post-HF methods including the implementation of analytic nuclear gradients for arbitrary boundary conditions. As already indicated above, particular emphasize is on HF and MP2 methods [6], which are a prerequisite for simulations based on RPA [7], hybrid and double-hybrid DFT schemes [8], as well as GW [9]. The four-center two-electron repulsion integrals (ERI), which in Mulliken notation reads as

$$(\mu\nu|\lambda\sigma) = \int d\mathbf{r}_1 \int d\mathbf{r}_2 \, \phi_\mu^*(\mathbf{r}_1)\phi_\nu(\mathbf{r}_1) \frac{1}{|\mathbf{r}_1 - \mathbf{r}_2|} \phi_\lambda^*(\mathbf{r}_2)\phi_\sigma(\mathbf{r}_2) \tag{1}$$

are of central importance for all wavefunction-based post-HF methods. In addition to well established integral screening techniques based on the Schwarz inequality

$$|(\mu\nu|\lambda\sigma)| \leq |(\mu\nu|\mu\nu)|^{1/2} |(\lambda\sigma|\lambda\sigma)|^{1/2} \tag{2}$$

to reduce the scaling from $\mathcal{O}(N^4)$ to $\mathcal{O}(N^2)$, a similar density matrix screening can also be employed to eventually sustain linear scaling with respect to the system size $N$. In the latter, the largest density matrix element $P_{\max} = \max\{|P_{\mu\lambda}|, |P_{\mu\sigma}|, |P_{\nu\lambda}|, |P_{\nu\sigma}|\}$ is used to screen elements smaller than $\epsilon_{\text{Schwarz}}$ via

$$P_{\max} \times |(\mu\nu|\mu\nu)|^{1/2} |(\lambda\sigma|\lambda\sigma)|^{1/2} \leq \epsilon_{\text{Schwarz}}, \tag{3}$$

where $P_{\max}$ is either the density matrix from the previous self-consistent field iteration, or from a converged semi-local DFT calculation, but ideally the propagated density matrix via second-generation Car-Parrinello AIMD of the previous timestep [2]. Yet, at the core of all implemented post-HF approaches are either the auxiliary density matrix method (ADMM) [6], or the resolution of identity (RI) approach [7]. The key ingredient of the former is the use of an auxiliary density matrix $\hat{\mathbf{P}}$, which approximates the original density matrix $\mathbf{P}$, but is substantially easier to compute due to being smaller in size, or more rapidly decaying than $\mathbf{P}$. For the case of computing the Hartree-Fock exchange (HFX), the exact energy $E_X^{\text{HFX}}[\mathbf{P}]$ is replaced by the computationally superior expression $E_X^{\text{HFX}}[\hat{\mathbf{P}}]$, whereas the difference between the two terms is corrected approximately at the semi-local DFT level. Hence,

$$\begin{aligned} E_X^{\text{HFX}}[\mathbf{P}] &= E_X^{\text{HFX}}[\hat{\mathbf{P}}] + \left( E_X^{\text{HFX}}[\mathbf{P}] - E_X^{\text{HFX}}[\hat{\mathbf{P}}] \right) \\ &\approx E_X^{\text{HFX}}[\hat{\mathbf{P}}] + \left( E_X^{\text{DFT}}[\mathbf{P}] - E_X^{\text{DFT}}[\hat{\mathbf{P}}] \right), \end{aligned} \quad (4)$$

where $E_X^{\text{DFT}}[\mathbf{P}]$ and $E_X^{\text{DFT}}[\hat{\mathbf{P}}]$ are the DFT exchange energies for the primary and auxiliary density matrices, respectively. The RI approximation, however, allows to substitute the computationally demanding four-center ERIs by just 2- and 3-center integrals by factorizing them via

$$(ia|jb)_{\text{RI}} = \sum_{PQ} (ia|P)(P|Q)^{-1}(Q|jb), \quad (5)$$

where $(P|Q)^{-1}$ is the inverse of the Coulomb metric over auxiliary Gaussian basis functions, i.e.

$$(P|Q) = \int d\mathbf{r}_1 \int d\mathbf{r}_2 \, \phi_P(\mathbf{r}_1) \frac{1}{|\mathbf{r}_1 - \mathbf{r}_2|} \phi_Q(\mathbf{r}_2). \quad (6)$$

Since the latter is a positive definite matrix, its inverse can be efficiently obtained by means of the Cholesky decomposition

$$(P|Q) = \sum_R L_{PR} L_{RQ}^T \quad (7)$$

followed by an inversion of the triangular matrix $\mathbf{L}$, i.e.

$$(P|Q)^{-1} = \sum_R L_{PR}^{-T} L_{RQ}^{-1}. \quad (8)$$

In this way, the factorization of the integrals $(ia|jb)$ can be written in compact form as

$$(ia|jb)_{\text{RI}} = \sum_P B_P^{ia} B_P^{jb}, \quad (9)$$

where

$$B_P^{ia} = \sum_R (ia|R) L_{PR}^{-1}. \quad (10)$$

Therein, the three-center integrals $(ia|R)$ are computed starting from integrals over atomic orbitals $(\mu\nu|R)$, so that the final expression for the elements $B_P^{ia}$ reads as

$$(ia|P) = \sum_\nu C_{\nu a} \sum_\mu C_{\mu i} \sum_R (\mu\nu|R) L_{PR}^{-1}, \quad (11)$$

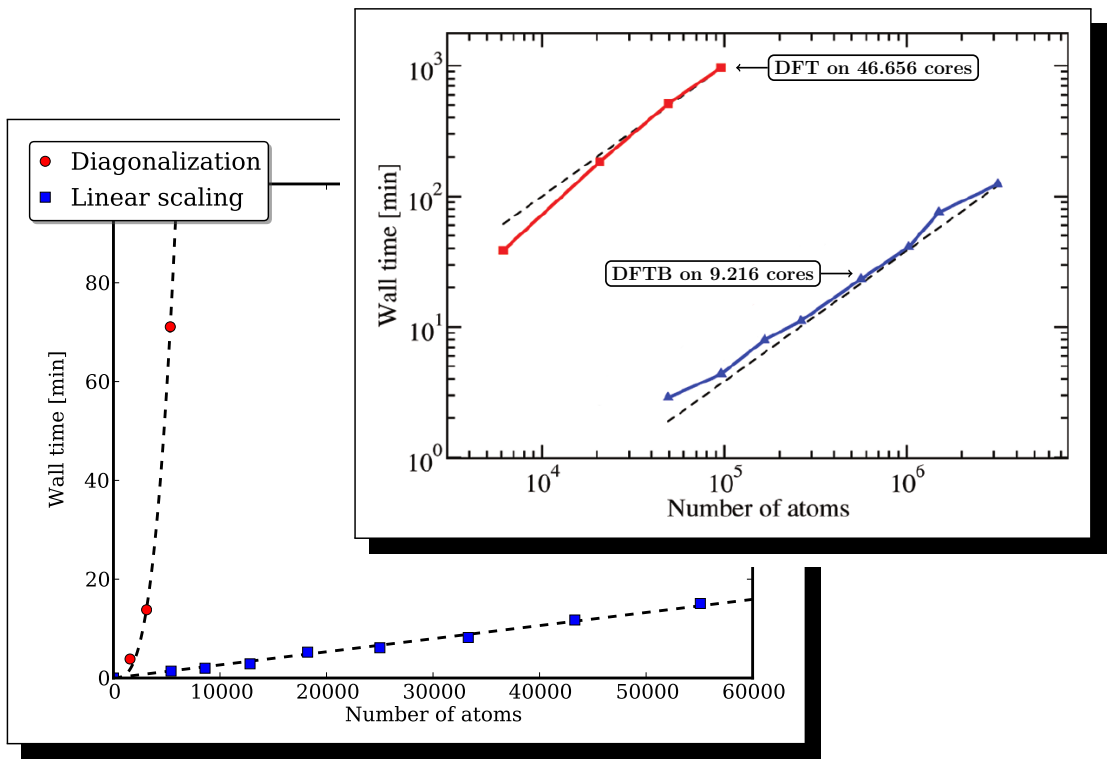

**Figure 1.** Total walltime for a single AIMD step for bulk water at ambient conditions as a function of system size on a Cray XT5.

where $\mathbf{C}$ is the molecular orbital coefficient matrix, i.e. $\mathbf{C}\mathbf{C}^T = \mathbf{P}$.

Another major development direction is the design of novel linear-scaling algorithms [10]. Within CP2K, the relation

$$\text{sign}\begin{pmatrix} \mathbf{0} & \mathbf{A} \\ \mathbf{I} & \mathbf{0} \end{pmatrix} = \text{sign}\begin{pmatrix} \mathbf{0} & \mathbf{A}^{1/2} \\ \mathbf{A}^{-1/2} & \mathbf{0} \end{pmatrix} \quad (12)$$

is employed together with various iterative methods to compute the matrix sign function

$$\text{sign}(\mathbf{A}) = \mathbf{A}\left(\mathbf{A}^2\right)^{-1/2} \quad (13)$$

to yield the inverses and (inverse) square roots of large sparse matrices $\mathbf{A}$ with a computational effort that scales just linearly with system size. Most importantly, however, the sign function can also be used for the purification of the Kohn-Sham matrix $\mathbf{H}_{\text{KS}}$ into $\mathbf{P}$ by the use of

$$\mathbf{P} = \frac{1}{2}\left(\mathbf{I} - \text{sign}\left(\mathbf{S}^{-1}\mathbf{H}_{\text{KS}} - \mu\mathbf{I}\right)\right)\mathbf{S}^{-1}. \quad (14)$$

The superior efficiency and linear-scaling potential of this approach has been demonstrated at the DFT level using a double-$\zeta$ valence polarization basis set, for up to a million of atoms in a massively parallel implementation, as shown in Fig. 1.

## 3. Meeting the Exascale Challenges

Standing at the brink of the exascale era, we can see that fundamental disruption in programming models and processor technology that were predicted a decade ago did not happen. From the developer's point of view, three trends originating in the petascale area are continued and emphasized: 1) the need of massive parallelism in terms of node and thread counts, 2) an even more widespread use of GPUs, and 3) the availability of mixed-precision matrix or tensor operation hardware accelerators.

CP2K is fundamentally well-suited to scale to immense levels of parallelism because it was designed as a massively parallel MPI application right from the outset. With the advent of multi-core processors, OpenMP directives were added to important loops, while leaving the underlying data layout unchanged to support multi-threading and hybrid parallelism. With that many concurrent threads the data has to be partitioned to prevent bottlenecks from reductions or atomic access collisions. The new library for Distributed Block-sparse Matrices (DBM) is a first step in this direction. It uses a fixed assignment of matrix block rows to threads, which eliminates the need for synchronization from most operations. We plan to refactor other primitives in the same way, in particular the grid data structures that power methods like GPW have a lot to gain from per-thread partitioning.

The modular structure and use of modern Fortran 2008 allowed the addition of GPU support at the level of GPU-accelerated libraries (e.g. DBM, COSMA, SpFFT, SPLA, grid, pw and Sirius), some of which were also spun off from CP2K as stand-alone libraries such as e.g. DBCSR. The challenge for the exascale era is the evolution and increasing diversity of GPU architectures. In the early days we mostly struggled with finding the sweet spot within the tight constrains set by small register files and scarce shared memory. In present systems, the bottleneck has now shifted to the PCI bus, where we are often limited by host-to-device communication. The way forward is to use GPU-aware MPI, which we are currently adding to DBM and will later roll our to other parts of the code. The status and results of these efforts can be tracked on the CP2K GPU dashboard at `https://www.cp2k.org/gpu`.

Lastly, we expect that the matrix and tensor processing units will have a profound and long-lasting effect on method and code development. These computing elements can achieve one to two order of magnitude more FLOPs for dense linear algebra operations in reduced or mixed precision, e.g. FP16 operations with FP32 accumulation. To exploit this potential, it is key to develop methods that heavily rely on dense linear algebra on medium sized local matrices, whose numerical inaccuracies due to mixed- and low-precision arithmetic can be rigorously compensated by the design of a modified Langevin-type equation [2].

In recent work, we have developed the submatrix method that is specifically designed with these design principles in mind. The core idea, as illustrated in Fig. 2, is to convert the evaluation of a matrix function on a large distributed sparse matrix into a large-scale parallel evaluation of the matrix function on many dense, but much smaller matrices. In combination with the purification scheme of Eq. 14, we were able to perform record sized linear-scaling electronic structure computations on systems with more than 100 million atoms, thereby achieving a sustained performance of 324 PFLOPs with an efficiency of more than 67% [10].

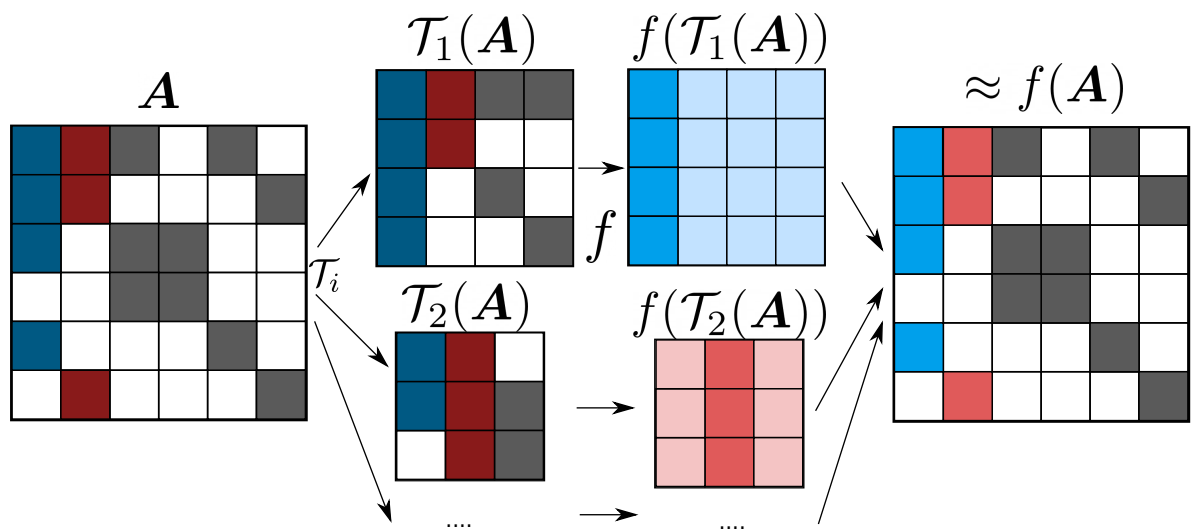

**Figure 2.** Schematic representation of the steps of the submatrix method for the approximate calculation of a matrix function $f(\mathbf{A})$ of a large sparse matrix $\mathbf{A}$. The first step is the construction of a submatrix $\mathcal{T}_i(\mathbf{A})$ for every column of the matrix $\mathbf{A}$. Then the matrix function is applied to the dense submatrices, i.e. $f(\mathcal{T}_i(\mathbf{A}))$, and finally the relevant result columns are inserted into the sparse result matrix. Figure reproduced from [10] under CC-BY.

## 4. Concluding Remarks

We conclude by noting that beside exploiting upcoming exascale supercomputers to facilitate ever more complex and accurate simulations, in the future alternative approaches such as deep neural networks (DNN), as well as quantum computing algorithms will find their application within CP2K. On the one hand this can be so-called learning on-the-fly schemes, where the decision if the calculation of an observable can be outsourced to an already sufficiently accurate DNN or needs to be conducted explicitly, which can then be used as training data to directly relearn the DNN, is done during the simulation itself. On the other hand, all sort of hybrid approaches can be imagined, where the majority of a particular quantity is explicitly calculated using a computationally simple, but approximate electronic structure method, which is augmented by a correction term that is represented by a DNN for instance [11]. De facto exact configuration interaction, or reduced density-matrix functional (RDMF) theory simulations will become feasible by the usage of hybrid quantum-classical algorithms in which the quantum mechanical expectation values of the RDMF are evaluated on a quantum computer, whereas the parameters of the trial states are optimized by a Car-Parrinello-like constrained minimization scheme within CP2K on a classical computer

[12].

## 5. Acknowledgements

The authors would like to thank the whole CP2K development team, who had been contributing to the code over the last two decades.

# 5 – DFT-FE Code: Systematically convergent large-scale DFT calculations using finite-element basis

Sambit Das [(1)], Phani Motamarri [(2)], Vikram Gavini [(1)]
[(1)] University of Michigan, Ann Arbor
[(2)] Indian Institute of Science, Bangalore

**Background and Current Status**

DFT-FE [1] is a recently developed open-source code (current v1.0 released in 2022) for conducting massively parallel, fast, efficient, and accurate density functional theory (DFT) calculations on hybrid CPU-GPU architectures. DFT-FE is based on a real-space formalism and adaptive higher-order finite-element (FE) discretization, providing systematic convergence with an ability to accommodate periodic, non-periodic and semi-periodic boundary conditions [2,3,4]. While we anticipate DFT-FE to be primarily used for pseudopotential calculations, the adaptive spatial resolution afforded by the FE basis also allows for systematically convergent all-electron calculations and, further, mixed all-electron and pseudopotential calculations [5].

DFT-FE employs a computationally efficient solver, namely the Chebyshev filtered subspace iteration approach [6], and a scalable implementation relying on the FE framework and mixed precision algorithms that increases the arithmetic intensity by reducing data movement costs on many-core and hybrid CPU-GPU architectures [4,7]. Consequently, DFT-FE exhibits close to quadratic scaling until 30,000 – 40,000 electrons, significantly delaying the onset of cubic computational complexity.

DFT-FE offers excellent parallel scalability on evolving heterogeneous architectures (cf. Figure 1), thereby enabling large-scale pseudopotential DFT calculations on tens of thousands of electrons at modest computational costs and low wall-times. In particular, DFT-FE has been executed on massively parallel many-core CPU (up to ~200,000 cores) and hybrid CPU-GPU architectures (up to 22,800 GPUs) with systems sizes reaching up to ~100,000 electrons. Notably, full ground-state calculations[1] involving 5,000-15,000 electrons can be completed in wall-times of ~1-3 mins on hybrid CPU-GPU architectures [4]. Further, benchmark ground-state DFT calculations involving 60,000 - 100,000 electrons have been demonstrated with wall-times of ~30-60 mins. The performance of DFT-FE reaching 46 PFLOPs (OLCF Summit supercomputer) on a metallic dislocation system comprising of ~100,000 electrons was nominated as a finalist for the 2019 ACM Gordon bell prize [7]. Overall, we anticipate that the capability of DFT-FE to conduct fast and systematically convergent DFT calculations on large-scale materials systems can aid computational studies in several fields, including applied physics, chemical sciences, materials science and metallurgy.

---

[1] All references to benchmark calculations in this section refers to accuracy levels of ~0.1 mHa/atom in energy, ~0.1 mHa/Bohr in force and ~5e-06 Ha/Bohr$^3$ in stress

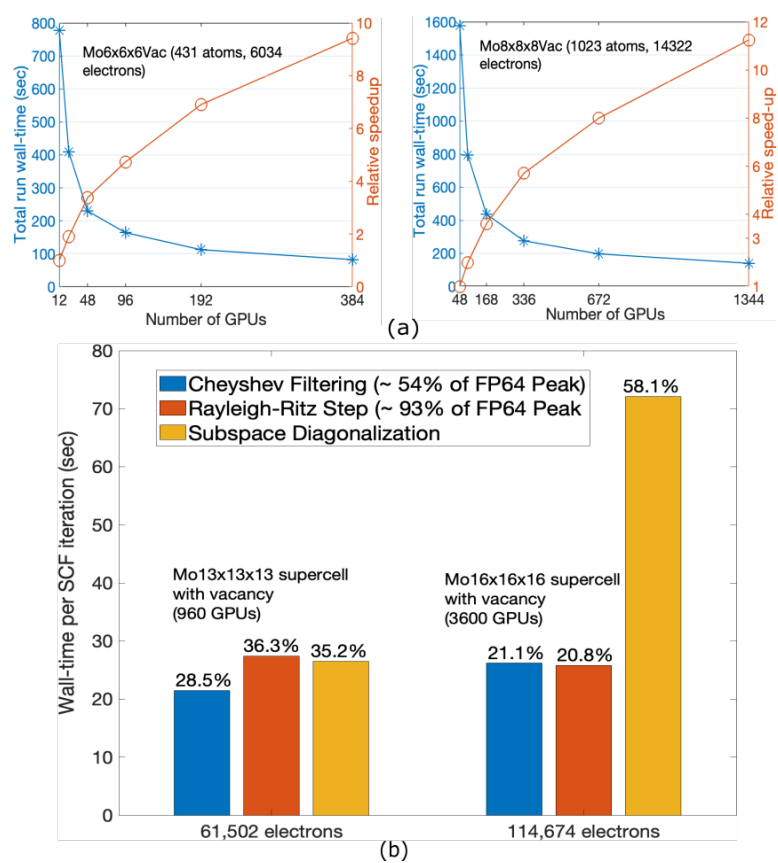

**Figure 1:** Performance profile of DFT-FE on the OLCF Summit GPU supercomputer. a) Strong parallel scaling of total ground-state run-time (including initialization and ionic force computation) for benchmark metallic systems comprising of 6,034 and 14,322 electrons. b) Breakdown of single SCF wall-time of Chebyshev filtered subspace iteration approach, involving Chebyshev filtering, Rayleigh-Ritz step (projection and subspace rotation) and subspace diagonalization, for large metallic benchmark systems.

### Development Priorities

We envision DFT-FE can be useful in tackling important challenges in a number of key areas, including: (a) S*tructural Materials*: For instance, understanding the behaviour of extended defects in metallic alloys and their connection to mechanical properties, such as strength and ductility, which requires fast, accurate and large scale DFT calculations reaching ~10,000 atoms; (b) *Energy Materials*: Such as understanding energetics and kinetics of solid-liquid and solid-solid interfaces, which require fast and large-scale ab-initio molecular dynamics (AIMD) and nudged elastic band (NEB) calculations on system sizes reaching a few thousands of atoms. These calculations would also require advanced exchange-correlation (XC) functionals—such as meta-GGA, Hubbard corrections—to model systems that exhibit strong correlations; (c) *Chemical Sciences* (catalysis) and *Biological Sciences* (understanding charge transport mechanisms), where accurate large-scale DFT calculations with improved XC functionals can provide essential insights and guidance.

In view of the above scientific applications and others, DFT-FE's developmental priorities over the next five years are broadly categorized into the following four areas:

(a) *Basis sets*: We plan to expand DFT-FE to include the enriched FE basis—a mixed basis, which comprises of the FE basis along with enrichment functions—that has been demonstrated to be computationally efficient for systematically convergent all-electron calculations [8]. We anticipate that the enriched FE basis can also further improve the performance of pseudopotential calculations.

(b) *Pseudopotential/All-electron functionalities*: We are implementing Projector Augmented Wave (PAW) formalism into DFT-FE that is expected to significantly reduce the FE basis functions by 3-10 fold in comparison to the norm-conserving ONCV pseudopotentials, presently supported by DFT-FE. We also plan to make the mixed pseudopotential all-electron capability—where some atom types are treated using all-electron description, while others are treated using a pseudopotential approximation—a common feature of the code, which can aid studies on spin defects in solids, among other applications.

(c) *Advanced XC functionals/Physics*: We are expanding the present capability of DFT-FE from semi-local functionals (LDA, GGA) to include the other widely used functionals. We are currently developing tensor-structured techniques for hybrid XC functionals to reduce their significant computational cost relative to semi-local functionals. In the near future, we also plan to implement other XC functionals such as meta GGA, Hubbard corrections, vdW corrections, and extend the code to account for spin-orbit coupling that is central to magnetism and spintronics. Further, for all-electron calculations, we plan to extend the present non-relativistic treatment to all-electron relativistic approximations, such as ZORA. As we expand the capabilities of DFT-FE, the focus will be on accuracy and scalability to enable robust, fast, and accurate calculations.

(d) *Capabilities related to structural relaxation and ion dynamics*: We are developing extended Lagrangian techniques for AIMD calculations that avoid the self-consistent field iteration in each MD step. We will further extend these techniques to advanced XC functionals. We anticipate that this will enable fast AIMD simulations reaching ~100 ps on system sizes containing a few thousands of atoms that can be crucial to understanding the complex structural and chemical processes, such as electrode-water interfaces in photoelectrochemical cells, ion diffusivity in solid-state electrolytes, among other applications. We also plan to develop and integrate on-the-fly machine learning strategies to achieve further acceleration in AIMD, NEB and structural relaxation simulations.

In addition to the above core developmental priorities, other development priorities of the DFT-FE team include: (a) Capability to conduct efficient and accurate inverse DFT calculations [9], which will leverage systematically convergent all-electron calculations in DFT-FE, to compute exact exchange-correlation potentials from many-body ground-state electron-densities. The resulting XC potentials can be useful to improve the existing XC functional models, as well as aid the development of accurate and transferable machine-learned XC (ML-XC) models, thus enhancing the model level accuracy of DFT calculations; (b) Capability to conduct fast and large-scale real-time time-dependent DFT calculations (TDDFT) [10]; (c) Implementation of post-DFT techniques including polarizability calculations and GW method to study electronic and optical properties in materials.

**Meeting the Exascale Challenges**

The current implementation strategies employed in DFT-FE are based on the HPC-centric philosophy of reduced data movement, including communication costs, leading to high throughput performance (cf. Figure 2(a)). This has been accomplished by using mixed precision arithmetic and employing asynchronous compute-communication paradigms in the Chebyshev filtered subspace iteration approach (ChFSI). These implementation procedures have resulted in excellent parallel scalability of DFT-FE demonstrated on up to 3800 hybrid CPU-GPU Nodes (22,800 GPUs) on system sizes involving ~100,000 electrons. Thus, DFT-FE is currently able to take advantage of existing pre-exascale architectures.

The performance benchmark calculations show that the subspace diagonalization cost in Chebyshev filtered subspace approach accounts for ~35-60% of the total SCF cost for large systems beyond 60,000 electrons (cf. Figure 1(b)). Currently, the subspace diagonalization is performed using a direct diagonalization approach available in the ELPA library. However, alternate approaches, such as Fermi-operator expansion, that rely on parallel dense matrix-matrix multiplications to evaluate the subspace projected density matrix can provide further scalability of the DFT-FE code making it well suited for extreme-scaling architectures. Furthermore, there is potential to exploit the tensor structured nature of the FE basis functions implemented in DFT-FE to develop a matrix-free computational framework that can accomplish on-the-fly FE discretized sparse matrix-multivector multiplications. These matrix-free techniques can further improve computational efficiency by reducing data access costs and floating-point operations. We anticipate that these strategies will enhance the computational performance (minimum wall times) of DFT-FE by a factor of 2- 4 fold.

The data-centric methodologies and implementation procedures will also be exploited in the following future methodological developments in DFT-FE:

(a) *Enriched FE basis and PAW framework*: HPC centric ideas of matrix-free or pseudo matrix-free techniques associated with FE discretized matrix times multi-vector operations reduce both floating-point operations and data movement costs. This can allow the use of higher-order FE basis polynomial degree in a computationally efficient manner while reducing the peak memory requirements, thus further enhancing the performance of DFT-FE.

(b) *Advanced XC functionals*: Implementation of tensor-structured techniques for hybrid functionals using batch tensor contractions and batch tensor reshape operations can result in efficient utilization of GPU compute and memory bandwidth. Further, nonlocal XC functionals like vdW-exchange or ML-XC can also benefit from using GPU accelerated tensor-structured techniques.

**Concluding Remarks**

We remark that DFT-FE is already able to efficiently utilize the many-core CPU as well as pre-exascale hybrid CPU-GPU supercomputers to enable fast, accurate and massively parallel DFT calculations on generic material systems reaching up to ~100,000 electrons. This is a result of the systematic convergence and locality of the adaptive higher-order FE basis, efficient and scalable Chebyshev filtered subspace iteration procedure, and HPC-centric implementation strategies on heterogenous computing architectures. Using the pre-exascale machines like OLCF Summit and NERSC Perlmutter, DFT-FE has recently been employed to study various large-scale science problems such as dislocations (cf. Figure 2(b)) and grain-boundaries in metallic alloys, spin defects that are promising candidates for spin qubits, electronic structure of large DNA molecules, and phase stability of doped thin-film ferroelectric materials.

Looking forward into the exascale era, we anticipate our proposed developmental priorities and hardware aware implementation strategies will further improve DFT-FE's core capabilities, which can enable routine application of DFT-FE to tackle outstanding scientific problems that require ab-initio calculations at larger length-scales and time-scales. Further, we emphasize that our proposed implementation of advanced XC functionals, ML-XC functionals, spin-orbit coupling, relativistic corrections, TDDFT and GW method will expand the range of materials systems and properties that can be studied using DFT-FE. Additionally, the C++ codebase of DFT-FE and our continuing emphasis on ensuring performance portability across a growing range of hardware accelerators and their programming languages (CUDA, HIP) will be critical for the long-term maintainability and extensibility of DFT-FE. Finally, we remark that we aspire to build upon the open-source credentials of DFT-FE to make DFT-FE into a community project with a growing base of developers and users, along with promoting active engagement between computational method developers and domain science experts.

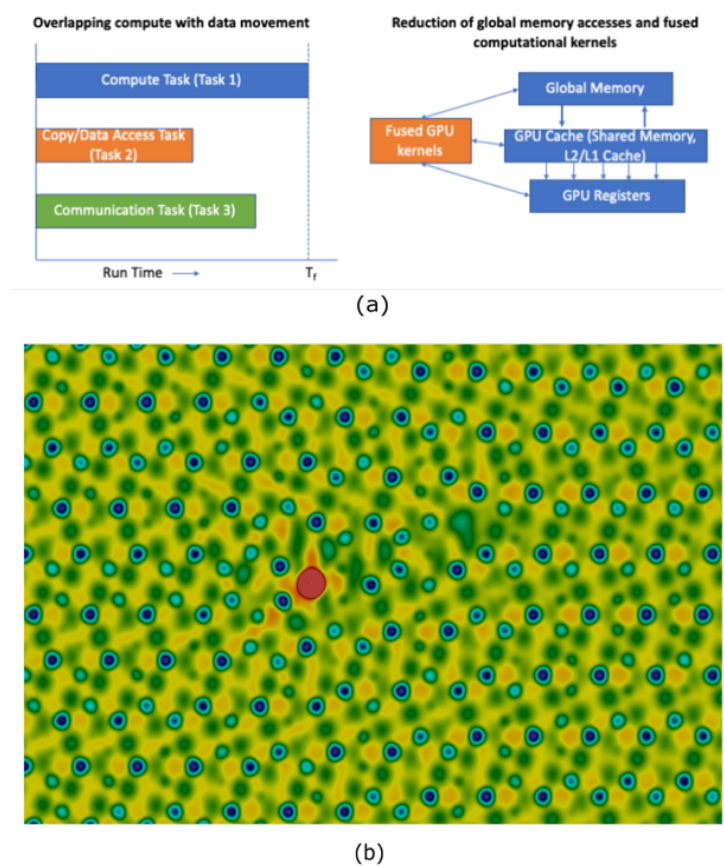

**Figure 2**: a) HPC-centric implementation strategies in DFT-FE on heterogeneous computing architectures. b) Electron-density contour of <c+a> pyramidal I screw dislocation in Magnesium with a Yttrium solute at the dislocation core. Computational domain contains ~6000 atoms.

**Acknowledgements**
DFT-FE has been the outcome of the efforts of many principal developers and contributors, and we refer to 'Authors' on the DFT-FE GitHub page [1] for the complete list of contributors. Various aspects of the code development efforts of DFT-FE are being supported by the U.S. Department of Energy Basic Energy Sciences, the Toyota Research Institute, and the Department of Science and Technology India. The algorithmic developments that form the core of the DFT-FE code have been supported by the U.S. Air Force Office of Scientific

Research and the U.S. Army Research Office. This work used the resources of the Oak Ridge Leadership Computational Facility, which is a DoE Office of Science User Facility supported by the Office of Science of the U.S. Department of Energy.

## 6 – An exciting Exascale Perspective

Andris Gulans[1], Alexander Buccheri[2], and Claudia Draxl[2]
[1]University of Latvia
[2]Humboldt-Universität zu Berlin

**Background and Current Status**

**exciting** [1] is a full-potential all-electron code, implementing density-functional theory (DFT), its time-dependent variant, TDDFT, and Green-function methodology, focussing on various excitations. The code employs (linearised) augmented plane waves and local orbitals (or loosely speaking just LAPW's) for valence and semicore electrons and explicitly treats core electrons via the radial Dirac equation. Since no shape approximation is required for describing wavefunctions, density or potential, this basis-set type provides a systematic path for reaching the complete-basis-set (CBS) limit, relying only on well-controlled numerical approximations.

Indeed, LAPW codes and specifically **exciting** serve as a reference tool for other approaches, especially those relying on pseudopotentials in ground-state as well as excited-state calculations. A remarkable example was the so-called ∆-test [2] that compared a sizable number of DFT codes regarding their performance for the equation of state on a benchmark set of 71 elemental crystals. Later it was shown [3] that **exciting** achieves microhartree precision for total energies in DFT calculations as demonstrated by comparison with multi-resolution analysis. Also in quasiparticle calculations, using the $G_0W_0$ approximation of many-body perturbation theory (MBPT), the CBS limit was attained without any extrapolation [4].

Importantly, with **exciting** one can study neutral excitations that appear in different kinds of spectroscopy, *e.g.*, optical absorption, electron-loss, EELS (electron energy loss spectroscopy), Raman, or RIXS (resonant inelastic x-ray spectroscopy). It employs the Bethe-Salpeter equation (BSE) to describe electron-hole excitations and provides means of analysis and visualisation. Figure 1 shows a hybrid charge-transfer exciton in pyrene@$MoS_2$ [6] as an example. Since LAPW calculations treat *all* electrons –also the low-lying ones– on equal footing, this method naturally gives access to core excitations and allows for an accurate description of excitations occurring from very low-lying up to shallow, more delocalised core states. **exciting** also enables explicit treatment of core-level spin-orbit coupling [5].

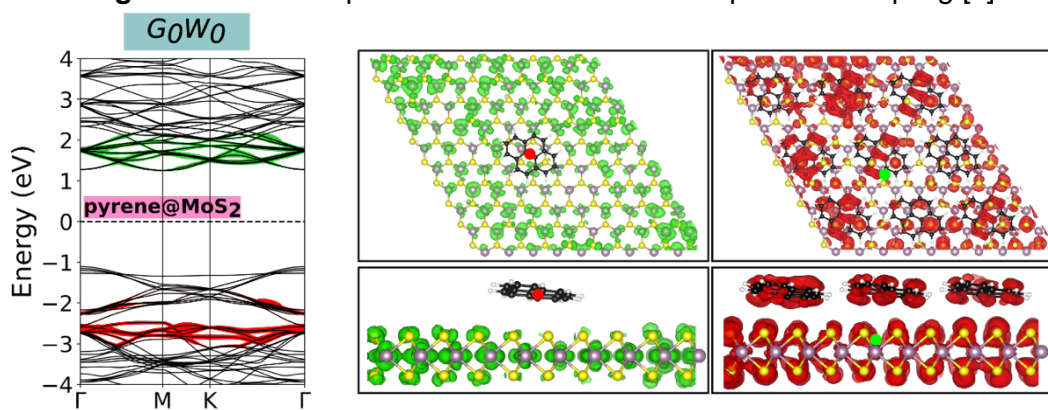

**Figure 1**. Reciprocal and real space representation of a charge-transfer exciton in pyrene@$MoS_2$. The red and green circles in band structure on the left indicate the contributions of the involved states. Their sizes are proportional to the transition weight. The real-space representations show the probability density of finding the hole of the electron-hole wave function given a fixed position of the electron (right panels) and vice versa (left panels). The electron (hole) probability distribution is depicted in green (red) with the corresponding hole (electron) position marked by the red (green) dots.

Such highly precise implementations should now be used to provide benchmark data to the community. This not only concerns small systems like the elemental crystals probed in Ref. [2] but should contain a representative set of materials from different classes and varying complexity. Moreover, various properties should be included, in particular those that may make potential discrepancies between different methods and approximations apparent. Such a task requires high-throughput calculations that ask for numerical efficiency and optimal performance on the exascale machines to come.

**Development Priorities**

High precision in LAPW comes along with high computational expenses. In part, it is due to the built-in complexity of LAPW compared to other formalisms. In part, it is due to the unavailability of low-scaling algorithms, be it generally or specifically concerning the LAPW basis. From a code-design perspective, an important aspect of current and future developments therefore concerns efforts to ensure that the wide variety of physical properties already available in the code can be computed not only for simple materials but also for complex systems to address highly topical research questions.

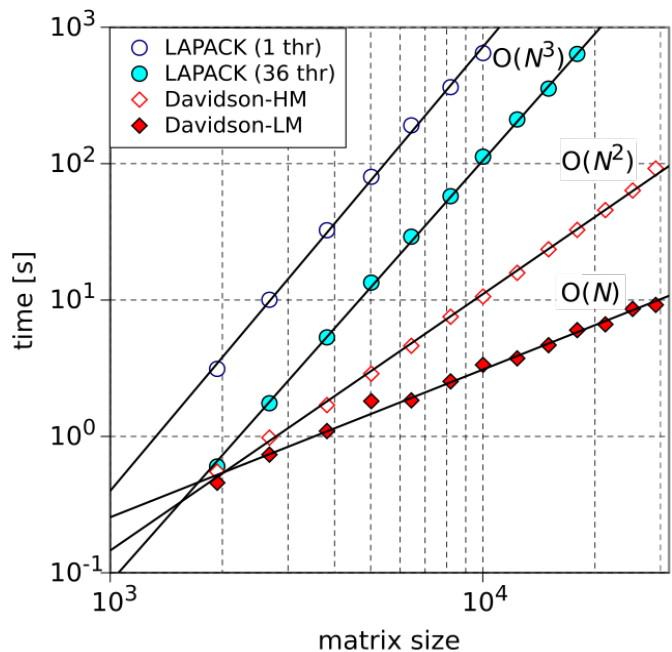

**Figure 2**. Performance of the iterative Davidson eigensolver in comparison to the standard LAPACK diagonalisation. The LAPACK timings are given for 1- and 36-thread calculations. The iterative eigensolvers with and without explicit matrix construction (labelled with HM and LM, respectively) were tested in single-thread runs.

For instance, ground-state DFT calculations require solving an eigenvalue problem; the optimal approach for low-dimensional and sparse systems is an iterative algorithm without explicitly constructing the Hamiltonian and overlap matrices. However, a high condition number is common in LAPW calculations, and iterative eigensolvers are then difficult to apply. To remedy this situation, we have modified the Davidson algorithm to make it applicable in these circumstances [3] and managed that way to reduce the computational and memory demand. With the example of the total energy of the CO molecule, Fig. 2 compares the performance of **exciting's** standard (LAPACK) and iterative (Davidson) algorithms for the diagonalisation. The size of the eigenproblem increases here with the simulation cell (amount of vacuum). We find that our implementation of the Davidson algorithm (without constructing the Hamiltonian explicitly) scales as O($N$) and outperforms the LAPACK eigensolver that scales as O($N^3$), yielding the same precision ($N$ being the number of basis functions).

For studying excited states, e.g. computing photoemission and optical absorption spectra, the *GW* approach and the solution of the BSE, respectively, remain the favourable methods, balancing accuracy and computational cost. **exciting's** all-electron $G_0W_0$ implementation formulates the key *GW* quantities –the dielectric function and the correlation self-energy– in terms of a product basis of Kohn-Sham wavefunctions. Computing the screening and the correlation self-energy in this manner scales as O($N^4$), and accounts for the majority of the time spent performing a $G_0W_0$ calculation. Analogously, the BSE represents the electron-hole wavefunctions as products of Kohn-Sham wavefunctions, and its conventional implementation scales as O($N^6$) as it computes all excitonic eigenstates. In both instances, we are striving for lower-scaling algorithms to improve performance. Furthermore, in addition to our mature BSE formulation, we are developing a novel approach, utilising real-space wavefunction interpolation [7] in order to treat systems where dense k-sampling is required to obtain convergence.

From the perspective of new scientific approaches, the **exciting** team is, for instance, actively researching the effect of phonon-mediated processes in excited state properties. Indeed, a first-principles treatment of electron-phonon interactions is essential in understanding temperature effects in both photo-electron and optical spectroscopy. The next release will feature a highly-parallelised implementation of density-functional perturbation theory (DFPT) in the linear-response regime. This will allow for a quantitative description of electron-phonon interaction in band-gap renormalisation, screening, and absorption processes, and form an essential ingredient in a higher-level theory of exciton-phonon coupling. All this goes hand in hand with approaches towards excitation dynamics with recent implementations of real-time TDDFT [8] and Ehrenfest dynamics [9].

**Meeting the Exascale Challenges**

**exciting** envisions reaching exascale performance via two distinct strategies. The first is the implementation of novel algorithms that facilitate optimal and massive parallelisation; the second is the automation of workflows to enable highly concurrent job submission and management.

In ground-state calculations of close-packed systems, the iterative Davidson eigensolver provides only little or even no gain over the conventional diagonalisation approach. To meet the challenge of performing large-scale all-electron calculations for materials science applications, we will rely on SIRIUS (https://github.com/electronic-structure/SIRIUS), a high-performance library for the LAPW basis, with asynchronous MPI and GPU support. The integration of SIRIUS in **exciting** enables the hand-off of the most computationally demanding parts, allowing one to perform ground-state calculations on systems up to an order of magnitude larger than currently possible.

Our aim is to exploit the massive potential for improving the parallel performance, as well as finding and implementing low-scaling algorithms for excited-state methods. In general, it is clear that migrating code from the traditional CPU programming model to support heterogeneous parallelism is a vital strategy in accelerating all aspects of electronic-structure calculations, and we stand to benefit greatly from this. To provide an example, within the NOMAD Centre of Excellence (https://nomad-coe.eu), a HORIZON 2020 project towards exascale computing, we are about boosting the efficiency of our $G_0W_0$ module. To ensure that we can tackle various system sizes most effectively, we pursue two targets: eliminating bottlenecks in the current implementation on the one hand, and taking a completely different route with a highly-parallelised implementation of the space-time method [10] on the other hand. The latter approach switches between real and momentum space representations for construction of the dielectric function, reducing the algorithmic complexity to $O(N^3)$.

A major effort in the NOMAD CoE goes to the development of libraries that cover major aspects of the involved algorithms. A part of them will be code-independent and thus can be shared among the partners that work with codes of different code families, *i.e.* different basis-set types. Other parts will be basis-set specific. This overall strategy is particularly important as exascale machines are not available in the European landscape yet; and those currently established rely on very different architectures. Such libraries can then be tuned for optimal performance on various hardware platforms. Most relevant for code developers and users, both types of libraries may finally be used also by other codes.

The last decade has seen substantial growth in molecular and materials databases, however in general, existing data on solids lack physical validation. Therefore, the second strategy of approaching the exascale concerns high-throughput calculations. We are developing automated end-to-end workflows to leverage **exciting's** high precision, such to provide benchmark data for various properties and material classes. Such data are also urgently needed in view of data-quality assessment in data collections [11]. This concerns particularly the aspect of *Interoperability* – the *I* in FAIR– when data from different sources are brought together such to employ them, for instance in machine-learning tasks. On-going work in this direction comprises various aspects that are key to reach our goals. These involve the automation of building transferable LAPW basis sets as well as the selection of optimal computational parameters. At the single-task level, automation is currently built within the Atomic Simulation Environment (https://wiki.fysik.dtu.dk/ase/). At the multi-task level, many are being composed to form complex, fully-automated workflows.

**Concluding Remarks**

**exciting** is tackling the exascale transition with a focus on theoretical spectroscopy, for both the valence and core region. A main goal is enabling calculations for complex systems that are both highly accurate (in terms of methodology) and highly precise (in terms of numerics). A major focus is thus put on the implementation of algorithms that can be optimally parallelised. The strategy is to collect these algorithms in libraries that can be tweaked on various (pre)exascale platforms and shared with other developers. The second route concerns the development of workflows to enable massive parallel job submission with automatised method- and material-specific setup and error handling. All this is embedded into the work carried out in the NOMAD CoE (see also Sections 1 and 7). Our work sets the stage for creating benchmark data for a wide variety of materials and properties as urgently needed for measuring the impact of methodology, approximations, and computational parameters on numerical results. But more than this, it will allow the community to explore exciting physics and tackle research questions that are not possible to answer with existing codes to date.

**Acknowledgements**

This work has received funding from the European Union's Horizon 2020 research and innovation programme through the NOMAD Center of Excellence, grant agreement No. 951786. We acknowledge partial support by the German Research Foundation (DFG) through the CRC 1404 (FONDA), project 414984028, and the NFDI consortium FAIRmat, project 460197019. We thank Ignacio Gonzales Oliva for providing Fig. 1.

# 7 – The FHI-aims Code: All-electron, *ab initio* materials simulations towards the exascale

Volker Blum[1], Mariana Rossi[2,3], Sebastian Kokott[4], and Matthias Scheffler[2]

[1] Duke University, Durham, NC, USA
[2] The NOMAD Laboratory at the FHI of the Max-Planck-Gesellschaft and IRIS Adlershof of the Humboldt Universität zu Berlin
[3] Max-Planck-Institute for the Structure and Dynamics of Matter, Hamburg, Germany
[4] Materials Simulations from First Principles e.V. (MS1P), Berlin, Germany

**Background and Current Status**

FHI-aims is a quantum mechanics software package based on numeric atom-centered orbitals (NAOs) with broad capabilities for all-electron electronic-structure calculations and *ab initio* molecular dynamics. It also connects to workflows for multi-scale and artificial intelligence modeling.

Since its foundation in 2004, the FHI-aims code has been designed with a clear set of goals. It should be numerically precise across the periodic table. It should be "all-electron" (not pseudopotential) and handle periodic systems (i.e., extended models of solids, surfaces, and nanostructures) as well as non-periodic systems (i.e., molecules and clusters). The code should support density-functional theory (DFT) with all relevant exchange-correlation functionals, and it should be amenable to correlated methods beyond DFT, i.e. the random-phase approximation and many-body perturbation theory (e.g., *GW*) based on Green's functions and the screened Coulomb interaction, as well as wave-function based correlation methods from quantum chemistry (MP2 and coupled-cluster theories). Furthermore, the code should scale efficiently from small to very large simulation sizes (thousands of atoms or more) and work seamlessly from limited hardware (laptops) up to the most powerful supercomputers available now or in the future. From the beginnings in 2004, the team grew to include several further key contributors by the time it was first released in 2009 [1]. Today, FHI-aims is a worldwide community project created by well over 150 individual contributors (https://FHI-aims.org/who-we-are) including support for key open source developments such as the ELPA library [2,3] for massively parallel eigenvalue solutions, the ELSI infrastructure for lower-scaling solutions [4], CECAM's electronic structure library [5], environments such as i-PI [6], FHI-vibes [7], the atomic simulation environment ASE [8], the open-source graphical interface for materials science (GIMS) [9], and many others (see Figure 1 and https://fhi-aims.org). The FHI-aims coordinators regularly organize schools and virtual tutorials (available at https://fhi-aims.org). Outreach efforts include industry, through the non-profit association MS1P (https://ms1p.org), ensuring that associated income is returned to the community via code advancements.

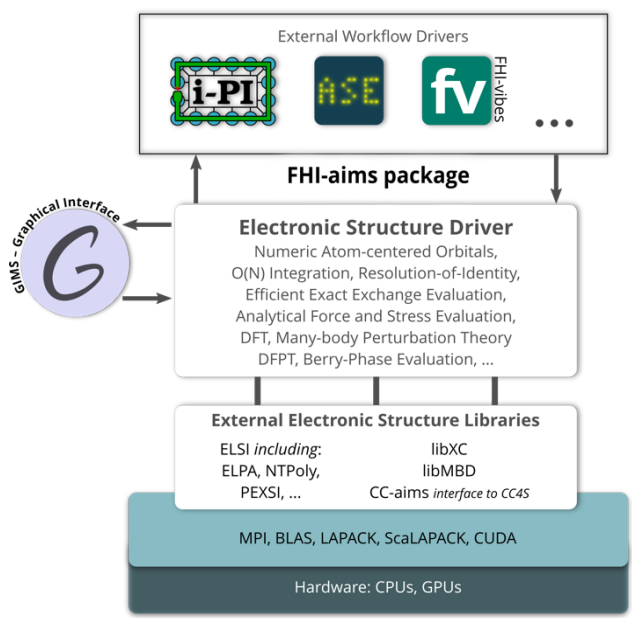

**Figure 1:** An overview of the FHI-aims code, including its core functionalities, external libraries directly coupled to the code, its integration with external workflow drivers and its integration with the graphical interface GIMS.

**Development Priorities**

The numerical foundation on NAO basis sets lies at the core of FHI-aims, allowing to represent the electronic structure of any problem in chemistry or materials science and engineering, without shape approximations. Support for and compatibility with Gaussian-type and Slater-type orbitals is contained in the code and important for excited-state calculations and electron-electron correlation beyond DFT. Key priorities that drive the ongoing developments include:

- *Usability*. Like many of its peer codes, FHI-aims is usable as a single binary at the command line of a terminal, through queueing systems at supercomputer facilities, or embedded in an ecosystem of separately developed and/or customized scripted tools for higher-level tasks [6,7,8]. The input to FHI-aims itself is simple, requiring only two input files and a few minimal keyword additions to get started; several tutorials and a browser based graphical interface, "GIMS", [9] (Figure 2) are also available. A key ongoing challenge lies in the ever-evolving complexity of high-performance computer systems, especially for the demanding applications of current and urgent interest. In this context, we note that critical tools in high-performance computing, such as message-passing interface (MPI) libraries, compilers and numerical libraries are insufficiently standardized and can present a steep learning curve for newcomers to the field. Reducing this learning curve through tutorials, testing, dedicated code advances and infrastructure will remain an overarching priority for future FHI-aims developments, in particular targeting new accelerator models towards the exascale era (see below).

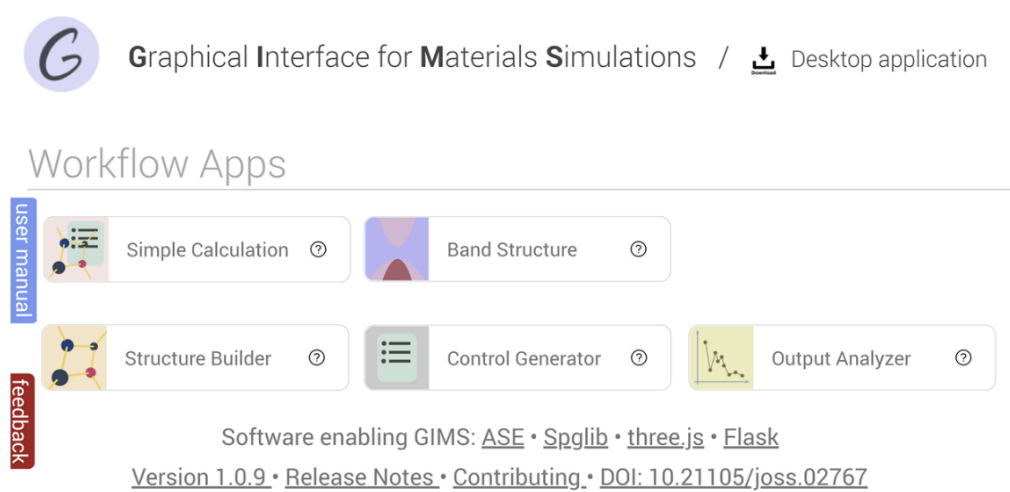

**Figure 2:** Start page of the Graphical Interface for Materials Simulations (GIMS) [9] for the FHI-aims code (print version). GIMS is completely browser based, i.e., immediately usable on any computer. The interface also supports the *exciting* code and, being built on the ASE [8], it is open to accommodate any other electronic structure code.

- *Community*. FHI-aims is a code based in a large academic community, especially when it comes to a plethora of new developments that a single, small team could not shoulder. Key examples are, e.g., refined density functionals that capture dispersion interactions accurately, real-time time dependent DFT, incorporation of nuclear quantum effects both in the code and by external tools, thermal and electrical transport calculations, *GW* approaches, and many more. Therefore, it is a matter of course to keep the FHI-aims code open, accessible and welcoming to a large community of existing and new users and developers.
- *Science*. In order to keep pace with the increasing needs of our field, continuous work on new features is essential. Examples of our ongoing work include efficient hybrid DFT for 10,000 atoms and more, relativistic formalisms capturing the full Dirac equation, important to capture spin-related phenomena, e.g., in "quantum materials", coupled-cluster theory for high accuracy of stability, reactions in and reactions between extended solids, and a plethora of approaches geared at accurately simulating excitations of the electronic and nuclear systems of a solids that connect to powerful spectroscopies as well as to device applications (e.g., optoelectronic or spintronic) by our experimental colleagues. Connecting the electronic structure foundation to artificial intelligence approaches in order to accelerate computational steps that do not need to repeated and/or can be predicted based on already existing information is a critical practical step for all these objectives.[10]

**Meeting the Exascale Challenges**

Many of our ongoing developments aim at enabling investigations of systems of higher complexity, systematic consideration of metastable states and temperature, and all this at significantly (urgently needed) higher accuracy than what is possible today. Importantly, the goal of utilizing ever-faster computing architectures goes beyond 'speeding up' state-of-the-art high-throughput computations that still employ the theory of the 1990s (through widely used, successful, but also fundamentally limited semilocal density functionals). Figure 3 shows the schematic reach of different levels of electronic structure theory; in FHI-aims, high-accuracy approaches to electronic structure theory are expected to benefit most directly from the exascale hardware.

Exascale architectures will be heterogeneous, featuring both CPUs as well as accelerators such as GPUs - the latter coming in various flavors (at the time of writing, at least NVidia, AMD, and Intel) and with different coding paradigms. Our strategy in FHI-aims has been to build the code around an "MPI first" paradigm, meaning that every computational method is foremost parallelized without any *a priori* restriction of execution across compute nodes in even the largest supercomputers (~200,000 cores were demonstrated already in 2011). Shared-memory parallelism within each node can be implemented through the MPI-3 standard for multicore processing systems where needed, but importantly controllable where needed from within the code. Through work for NVidia GPUs, we already have a working strategy to treat particular computational hotspots by GPU offloading [2,3,11]. Figure 4 shows the impressive power that can be leveraged for large problems on even a few nodes of the pre-exascale computer Summit (42 Power9 cores and six NVidia V100 GPUs per node) at Oak Ridge National Laboratory, compared to many tens of nodes of the Cori Intel Haswell system (32 cores per node) at National Energy Research Center only a few years earlier. Ongoing work focuses on extending this paradigm throughout the code as well as to the newer AMD and Intel architectures.

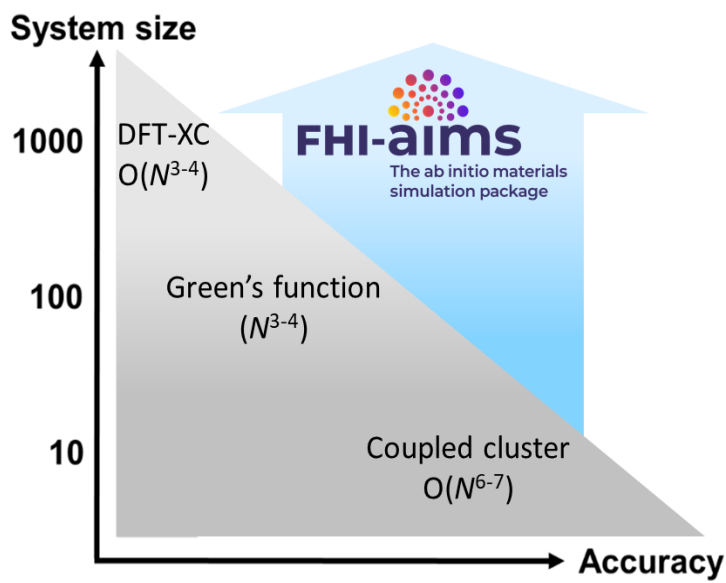

**Figure 3:** Schematic scaling of various electronic-structure methods and system size (# of atoms) currently possible. For DFT, we indicate cubic scaling since lower scaling is typically not yet reached for ~1,000 atoms in dense systems, in our experience. Direct electronic-structure calculations for mid- to large-scale systems at the highest levels will benefit most dramatically from successful exascale implementation. Our efforts in FHI-aims will focus especially on these highest levels, e.g., in work done in the NOMAD Center of Excellence (https://www.nomad-coe.eu).

FHI-aims already includes advanced exchange-correlation methodologies such as the random-phase approximation, second-order Møller-Plesset perturbation, and coupled-cluster theory. These are presently being extended for the exascale hardware. For instance, the scalability and performance of large-scale DFT calculations is determined by the eigensolver. For hybrid DFT calculations, a second bottleneck is the evaluation of the non-local exact-exchange part of the Fock matrix, and for *GW*, RPA, and CC calculations it is determined by algebraic tensor operations. We are tackling these challenges together with wider community efforts, e.g., ELPA [2,3], ELSI [4], and the NOMAD Center of Excellence (https://www.nomad-coe.eu) (see also Sections 1 and 6).

For large systems, the time spent for diagonalization in DFT is always a potential bottleneck. $O(N^3)$ ($N$: system size) scaling dense linear algebra approaches remain competitive with alternatives up to thousands of atoms in our benchmarks. We are therefore helping to enhance the eigensolver ELPA in terms of functionality, performance and energy efficiency, in order to deliver an exascale version. Through the ELSI infrastructure project,[5] we are also connected to other highly efficient solvers that scale lower than $O(N^3)$, such as NTPoly, $O(N)$, or the PEXSI solver, $O(N^2)$. Google's Tensor Processing Units (TPUs) were recently employed to accelerate FHI-aims' conventional DFT (no sparsity assumptions) to almost 250,000 orbitals.[12]

When non-local operators are needed (e.g., for hybrid functionals), the bottleneck is created by the formally (without accounting for sparsity) quartic scaling of the method. For hybrid DFT, $O(N)$ scaling has long been realized, but overhead remains especially for intermediate-sized systems. For the even more challenging beyond-DFT methods, we are working on providing low-scaling, efficiently load-balanced implementations for RPA, MP2, and *GW*, using the real-space and imaginary time treatment or variations thereof. This will include, most critically, sparse matrix-matrix operations, batched matrix-matrix multiplications, and data rearrangement.

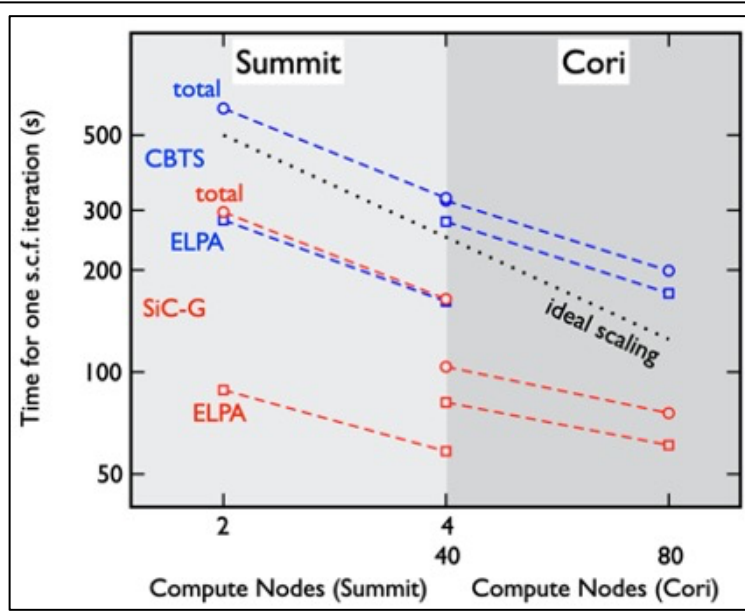

**Figure 4:** Visualization of computational time as a function of number of compute nodes used, for one self-consistent field (s.c.f) iteration of semilocal DFT for two large systems on the Cori supercomputer (CPU only, 40 and 80 nodes, right side) and on the Summit supercomputer (CPU+GPU, 2 and 4 nodes, left side), showing previously published data from Table 1 of Ref. [3]. FHI-aims' "light" settings were used. "CBTS" (blue curves) is a 3,000-atom periodic supercell model of a $Cu_2BaSnS_4$ semiconductor. "SiC-G" is a 3,376-atom slab model of a graphene layer on a SiC(111) substrate. Data is shown for the total time per s.c.f. iteration (circles) and for the portion consumed by the eigenvalue solver ELPA. On Summit, a portion of the s.c.f. cycle not related to the eigensolver (the electrostatic potential) is not yet accelerated on GPUs. Further details are provided in Ref. [3].

**Concluding Remarks**

FHI-aims is used and advanced by a great community. The code has been used for a wide range of calculations and, via workflows, many multi-scale modeling and artificial intelligence analyses (see e.g. Ref. 10). Pre-exascale architectures are already well supported. In addition to "heroic" largest-scale calculations, FHI-aims is also capable of launching an essentially unlimited number of separate, ensemble-parallel calculations at once via split MPI communicators, a mode of operation that is well suited for the exascale regime. Exascale computing may have a significant energy footprint. Here the FHI-aims community works on systematic optimization, e.g. by active learning strategies and workflows that start from the knowledge of the NOMAD data base (https://nomad-lab.eu/services/repo-arch) and make educated decisions for special DFT calculations in order to create a reliable and informative data pool for a faithful AI description. Statistical mechanics and multi-scale modeling require long time and length scales and here, for example, the hand-shake linkage to machine-learned potentials is being developed.

**Acknowledgements**

Foremost we thank the wide community of FHI-aims users and developers, the latter are listed at https://fhi-aims.org/who-we-are. We do not have space to list everyone individually, but we would like to particularly thank the Max Planck Computing and Data Facility (Garching, Germany), for their continued support. We thank Bernard Delley for several discussions that started well before the FHI-aims developments began and continued afterwards. We also thank Martin Fuchs for key discussions and software contributions during the initial development phase and Jörg Behler and Karsten Reuter for advice during the inception of FHI-aims. This work received funding from the European Union's Horizon 2020 Research and Innovation Programme (grant agreement No. 951786, the NOMAD CoE), and the ERC Advanced Grant TEC1P (No. 740233). Work on ELSI, ELPA-GPU and the CECAM ESL was partially supported by the National Science Foundation under Award Number 1450280.

# 8 – NWChemEx
Ryan M. Richard, Ames Lab and ISU
Theresa L. Windus, ISU and Ames Lab

**Background and Current Status**

Electronic structure theory (EST) attempts to accurately approximate solutions to the electronic Schrodinger equation in order to accurately predict chemical properties and reactions. Within the field of EST, NWChem[1] is arguably the de facto choice for high-performance molecular EST studies. Unfortunately, NWChem was initiated over 20 years ago and computer hardware and software architecture has far surpassed the original design. With the advent of the exascale era, the decision was made to focus efforts on NWChemEx (NWX),[2] a complete rewrite and redesign of NWChem from the ground up.

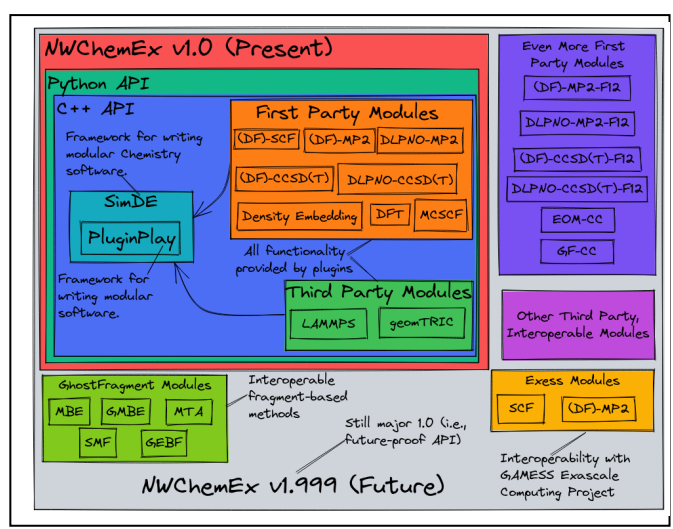

**Figure 1.** Top-left inner red box: Key design elements and anticipated functionality of NWX for version 1.0 (released on a rolling basis). Outer grey box: How we anticipate the NWX ecosystem to grow in the future. See text for details.

Figure 1 shows the key design points and features of NWX. The upper-left, red box lists the features anticipated to be in the version 1.0 release (tentatively slated for the end of calendar 2022). Like the original NWChem, NWX's primary niche is still high-performance EST. Unlike the original Fortran-based NWChem, the core of NWX is written using C++17 with nearly all functionality accessible via Python bindings and targeted for heterogeneous computers. NWX also has the fairly unique (at least from the perspective of EST) feature of being entirely plugin based. At the heart of NWX is PluginPlay,[3] a generic C++ framework for runtime manipulations of a program's call graph. PluginPlay views the entire program as a series of interconnected modules where each module tends to have the granularity of a function in a conventional EST package. While PluginPlay modules necessarily have generic APIs, SimDE defines additional module APIs using familiar chemistry concepts (e.g., molecules, basis sets, wavefunctions). SimDE forms the basis for the remainder of NWX, which is implemented as a series of encapsulated modules. Ultimately, the reliance on modules aids in: performance-tuning, reuse of NWX modules by other EST packages, incorporation of third-party contributions into NWX, and rapid prototyping of new theories. NWX also strives to have performance portability by providing separation of concerns through high level APIs and algorithms that have specific implementations for CUDA, HIP and SYCL (for example) underneath. OpenMP could be supported in this approach, although it is currently not as performant.

**Development Priorities**

The NWX team has taken the need to rewrite the original NWChem package as an opportunity to create our ideal next-generation EST ecosystem. In particular, not only must NWX perform well, it must also adhere to modern computer science practices, be highly customizable and extensible, easily integrate into existing scientific workflows, be developer-friendly, and be as user-friendly as possible. Historically the EST community has viewed these design points as being at odds, making this arguably a tall order. One of the largest and most time-consuming challenges associated with the development of NWX, was the literal design of NWX.

Aside from the design of NWX, our other top development priorities have been implementing: the computational infrastructure underlying NWX (i.e., PluginPlay and SimDE), electronic embedding, density functional theory (DFT), and domain local, pair-natural orbital (DLPNO) based implementations of second-order Moller-Plesset perturbation theory (MP2) and coupled cluster with single, double, and perturbative triple excitations [CCSD(T)]. Multiconfiguration self-consistent field (MCSCF) will also be available to facilitate modeling systems with strong-correlation. While not a development priority, we have also implemented both conventional and density-fit (DF) versions of SCF, MP2, and CCSD(T). NWX version 1.0 will be able to compute energies for all of the aforementioned methods and gradients for SCF and DFT, at a computational cost which is reasonably competitive with existing EST packages. In addition, third-party software, such as LAMMPS[4] and geomeTRIC,[5] will enable molecular dynamics and optimizations.

Admittedly the road to NWX version 1.0 has been a long one fraught with challenges. Chief among these challenges has been the learning curve associated with transitioning to modern C++17 object-oriented programming, the quickly changing hardware/software landscape, and the difficulty in attracting new talent to the project. Nonetheless, given the prevalence of object-oriented programming in the greater programming community, and with much of the HPC community transitioning to C++, we feel that our effort will be justified and will pay-off in the long run. It is our opinion that NWX version 1.0 is an excellent research and development platform that will provide for future functionality and flexibility for developers and users alike.

Over the next five years there will be a bifurcation of our development efforts into first- and third-party priorities. The first-party priorities are associated with adding features to improve the accuracy and functionality of NWX. Initial implementations of plane-wave based DFT and classical molecular dynamics will be incorporated. Effective core potentials, relativistic effects, spectral properties, solvent methods, and excited state methods will be added since these are commonly required by NWChem users. Explicitly-correlated (-F12) implementations of DLPNO-MP2 and DLPNO-CCSD(T)[6] will be added to increase the fidelity of the linear-scaling methodologies.

NWX is really envisioned as part of a larger software ecosystem. In the ideal ecosystem the broader EST community develops functionality which adheres to community-wide standardized APIs. Any third-party software exposing these standard APIs, would then, via the module system, be interoperable with NWX. SimDE proposes a set of standard APIs. The last NWX development priority is to leverage the SimDE APIs to grow NWX's functionality. Our initial efforts will focus on GhostFragment, a package for fragment-based methods, and EXESS[7] (an Exascale Computing Project offshoot of the GAMESS[8] package). We also anticipate the incorporation of NWX into scientific workflows important for large campaigns and machine learning to solve new scientific challenges.

**Meeting the Exascale Challenges**

The modular nature of NWX is one of the keys to achieving high-performance on today's machines, and to continuing to achieve high-performance on future machines. More specifically, since every major step of NWX is implemented as a module, and since the modules are only coupled at runtime, modules which are found to be bottlenecks can be rewritten without breaking the surrounding code. In practice modules are relatively fine-grained, so this allows us to, for example, focus on porting the evaluation of the exchange-correlation potential to GPUs without having to rewrite the remainder of the DFT algorithm. Of course, in practice achieving high-performance requires porting more than just the exchange-correlation potential to GPU; however, with the separation of concerns imparted by the module system this can be done piecewise. This approach also allows flexibility for sandboxing new computational approaches and theoretical methods. If tighter coupling is required for performance, a "higher-level" module could also be developed. An additional advantage of the runtime coupling is that it makes reasoning about the code's logic much easier by reducing branching points, which in turn facilitates additional parallelism opportunities.

The other key to NWX's performance is the use of object-oriented programming. The vast majority of objects in NWX rely on the "Pointer to implementation" (PIMPL) idiom. The PIMPL idiom, combined with clever design, allows the API of an object to be largely decoupled from how the operations are actually implemented. For example, the parallel environment is an abstraction that can support multiple parallel paradigms such as MPI and/or threads on CPUs and GPUs. Another useful class provides general mechanisms for caching results for data reuse and restart - again with the potential of multiple solutions depending on the needs of the user or developer.

Our tensor class is a textbook example of the advantages of object-oriented programming and is summarized in Figure 2. The SimDE API defines only one tensor class: TensorWrapper. Under the hood TensorWrapper relies on the Allocator and Shape abstractions to implement details such as whether the tensor lives in memory, if the tensor is built on the fly, if the tensor is GPU-based, if the tensor is sparse, etc. These details are then currently mapped to the TiledArray[9] back-end. As an alternative tensor model we have also developed the Tensor Algebra for Many-body Methods (TAMM)[10] software. The beauty of using TensorWrapper for the SimDE API is that it makes interfacing software developed directly with TiledArray or TAMM possible.

The backend for the tensor is set during construction, after which point module developers have little need to worry about the backend again. The vast majority of modules take as inputs already created tensors and create new tensors by performing fundamental tensors operations (e.g., adding, contracting, slicing, or permuting) on the input tensors. The consequence of this API is that nearly all of the complexity of the data movement is encapsulated by the tensor class (and an expression template layer). Admittedly, this design makes the backend of the tensor class extremely complicated; however, this complexity is now localized and when a tensor scenario is optimized, those optimizations are immediately used throughout the code. Furthermore, the resulting physics is extremely easy to implement (e.g., our in-memory and direct SCF/MP2 implementations use the same code which looks like tensor equations), making rapid prototyping viable, and dramatically lowering the barrier for new contributions.

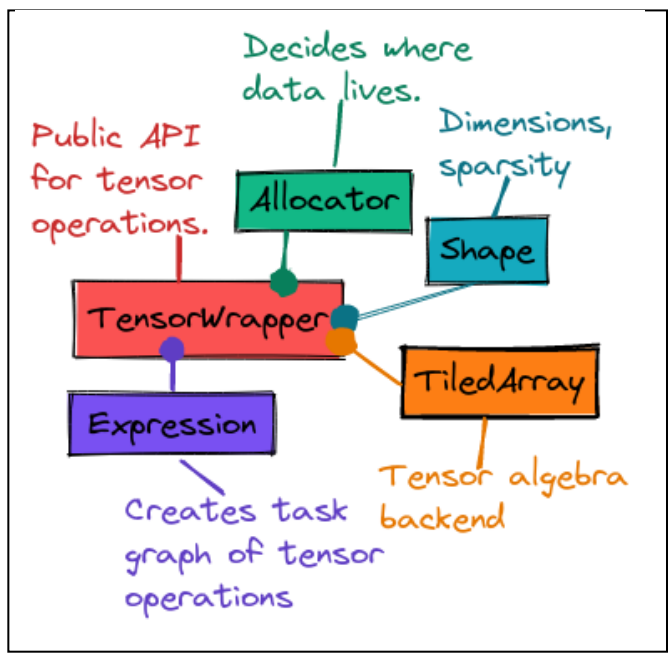

**Figure 2.** Brief overview of how tensors work in NWX. TensorWrapper is a layer built on top of the high-performance tensor library TiledArray. See text for details.

**Concluding Remarks**

A version 1.0 release of NWChemEx (NWX) is planned for the end of 2022. The release is bittersweet in that it is not as feature complete as we would like it to be, but at the same time we are very excited to show off what we think is the beginning of a new generation of electronic structure theory (EST) packages. In our opinion, basing NWX on PluginPlay results in an EST package that is a substantial departure from most traditional EST packages. In particular, the PluginPlay framework allows users/developers to customize (at runtime) just about every aspect of the code via the module system. Besides ensuring a modular package, PluginPlay also automates (to the extent possible) saving/loading EST calculations, generating documentation, and data archival. The modular approach along with separation of concerns also facilitates the use of high-level APIs with high performance implementations to ensure portable performance capabilities. After the initial release of NWX, rolling releases will add modules implementing conventional HF and MP2, as well as emerging linear-scaling variants of MP2, CCSD, and CCSD(T). The performance of these modules is currently competitive with implementations found in other EST packages, although only some of these modules will be fully exascale ready upon release. Future development priorities are focused on improving performance of existing features, implementing additional first-party modules (e.g., F12 methods and additional nuclear derivatives), and filling out NWX's feature set through third-party module support.

**Acknowledgements**

We thank the full NWChemEx development team and collaborators for all of their efforts in bringing NWChemEx to life. You have made this work possible. This research was supported

# 9 - PARSEC: Real-Space Pseudopotential Density Functional Theory Code

Kai-Hsin Liou[1], Mehmet Dogan[2], James R. Chelikowsky[1,2,3]
[1]McKetta Department of Chemical Engineering, University of Texas at Austin
[2]Center for Computational Materials, Oden Institute, University of Texas at Austin
[3]Department of Physics, University of Texas at Austin

**Background and Current Status**

PARSEC is a package for computing the electronic structure of materials. The code is based on real-space pseudopotentials constructed using Kohn–Sham density functional theory (DFT). Grid points represent the physical variables (Figure 1). A solution to the electronic structure problem provides us insight into the fundamental phenomena that occur in the microscopic world and can be used to predict materials properties.

The combination of real-space DFT and pseudopotential theory constitutes a powerful aid for scientists and engineers to search for high-performance materials. Created in 1994, PARSEC was the first practical code to solve Kohn–Sham problems on a real-space grid with the derivatives of physical quantities expanded using high-order finite differences [1]. Real-space methods possess numerous advantages for large-scale simulations [2, 3], with the avoidance of global communication from fast Fourier transformation being a primary enabler for superior scalability.

PARSEC was specifically designed for the electronic structure of nanostructures. For example, PARSEC often assumes a finite domain beyond which the wave function vanishes. As a result, PARSEC can handle defects or charged systems naturally and efficiently. PARSEC also supports periodic boundary conditions (1-D, 2-D, and 3-D), as well as spin polarization, spin-orbit interactions, and Born–Oppenheimer molecular dynamics simulations [4]. The output wave functions can be used as the input for further excited-state calculations (currently PARSEC provides interfaces to BerkeleyGW and NanoGW). PARSEC supports Troullier–Martins norm-conserving pseudopotentials and LDA/GGA functionals.

The eigensolver of PARSEC has changed over the years along with the development of high-performance computing. Eigensolvers are essential for enhancing the performance of electronic structure codes. In 2006, we proposed Chebyshev-filtered subspace iteration method (CheFSI)–an efficient algorithm tailored for real-space methods [5]. Based on the observation that during a self-consistent-field process the charge density and the wave functions are improving simultaneously, the inner (an eigenvalue problem) and outer (convergence of the potential) loops are fused. Upon convergence of the potential, the wave functions are a ground-state solution as well. CheFSI not only allows a fast solution for large systems, but creates a pathway to new research directions for leveraging the power of contemporary high-performance computing architectures.

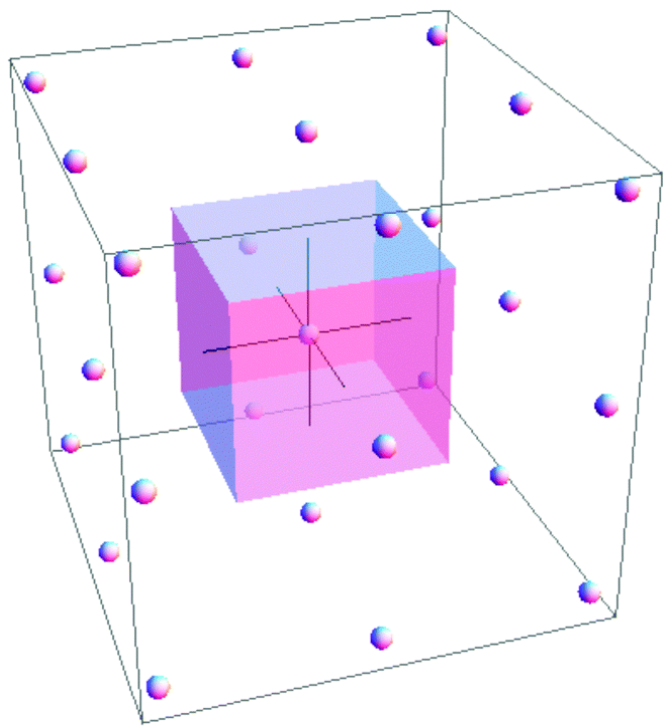

**Figure 1.** A cubic simulation domain with a regular real-space grid. The wave functions are defined on the grid points. Reproduced from Ref. [6] with permission from the PCCP Owner Societies.

**Development Priorities**

We follow three priorities in developmental activities: accuracy, ease of use and implementation, and performance.

**High-level exchange-correlation functionals.** The exchange-correlation functionals used in DFT are key to the inclusion of many-body effects. Proposed higher-level functionals, such as meta-, hybrid-GGA, and beyond, target a better capture of many-body interactions. However, higher-level functionals tend to result in a higher computational load. We plan to implement efficient hybrid functionals that are tailored for real-space methods in order to capture the physics of complex systems in an efficient and economical way.

**Support for more pseudopotential types.** Currently PARSEC supports Troullier–Martins norm-conserving pseudopotentials in various formats. Although our pseudopotentials database is comprehensive and suitable for most chemical environments, it is important to enable users to generate pseudopotentials by themselves to better meet their needs or use pseudopotentials from other databases. As a result, we may wish to support other pseudopotential formats.

**Efficient high-order forces.** The computation of accurate inter-atomic forces is crucial in geometry optimization and *ab-initio* molecular dynamics. In a real-space calculation, one can achieve more accurate forces by adopting a finer real-space grid. However, the computational cost notably increases as the grid spacing decreases. An ongoing activity seeks to improve the quality of forces by finite-difference methods without the use of unnecessarily fine grids. In 2015, we proposed an efficient way to compute accurate forces by high-order integration techniques [6]. We plan to continue this development. We also plan to implement variable-cell geometry optimization and the isothermal-isobaric ensemble for molecular dynamics. Combined with high-quality forces, these will be useful tools for materials science practitioners.

**Space-filling curves based grid partitioning for non-orthogonal lattices.** Space-filling curves (SFCs) can generate efficient real-space grid partitions and grid-point ordering [7]. However, our current implementation is for orthogonal lattices. While in many cases one can use an orthogonal cell instead of non-orthogonal ones, PARSEC should support an arbitrary cell shape and will extend the capability of SFCs-based grid partitioning algorithm to non-orthogonal cells. We note that different sizes or shapes of systems may require different SFCs for better performance. A study on optimal SFCs is important for exploiting the computing power of modern and future high-performance computing with massive vectorization processing units.

**Support for GPUs.** We will support the use of GPUs for various parts of the CheFSI algorithm to speed up calculations. The use of GPUs benefits efficient sparse matrix–vector multiplication (SpMV) as well as dense matrix operations (*e.g.*, large matrix–matrix multiplication and dense eigenvalue decomposition). Furthermore, Das *et al.* have demonstrated that GPUs can bring significant speedup for real-space finite-element methods [8], which confirms the potential of applying GPUs to the finite-difference methods.

**More efficient mixers.** Mixers are important in self-consistent DFT calculations. Halving the number of self-consistent-field iterations might be easier than making the eigensolvers run twice as fast (assuming the run time per iteration is comparable). Mixing algorithms that incorporate machine-learning techniques is a possible tack. As machine-learning methods become more efficient, one might capture information from wave functions, electron charge density, and potentials to better approximate the potential with few iterations.

**Meeting the Exascale Challenges**

A fast and scalable eigensolver with multilevel parallelization is central in addressing the exascale challenges. In electronic structure calculations, solving the eigenvalue problems is often the main bottleneck. To be flexible on the new accelerators, we plan to develop faster and better scalable eigensolvers by expanding and optimizing the current multilevel parallelization scheme in PARSEC. The topmost level is at the spectrum of the Hamiltonian, followed by Kohn–Sham states, real-space domain, grid blocks, and grid points in one grid block. The last three levels together are for SpMV, where GPUs could play an important role in accelerating calculations.

**Spectrum slicing.** Spectrum slicing provides a high-level decomposition of the problem [9, 10]. As the first parallelization level, the spectrum of interest is sliced and the eigenvalue problem is divided into sub-problems. The sub-problems can be solved simultaneously. Along with the problem, available processes are divided into groups–one for each slice. The processes of the same group focus on solving for the eigenpairs assigned to their slice and are completely independent of other slices. Communication between slices happens only when updating the electron charge density and potentials. We note that if there are k-points, multiple spins, and/or representations due to symmetry, there could be other layers of parallelization on top of that of spectrum slicing.

**Kohn–Sham states.** Each slice group solves a smaller eigenvalue problem using the CheFSI algorithm [5], in which the filtering step can be parallelized over the Kohn–Sham states.

Processes of the same slice group are divided into column groups and each column group performs filtering without communicating with other column groups.

**Domain partitions, grid blocks, and grid points.** The real-space grid is partitioned into sub-domains. Hilbert SFCs can be used in the partitioning to achieve efficient SpMV [7]. SpMV constitutes the filtering step, which renders itself the key to an efficient CheFSI algorithm. SpMV is performed by the processes of the same column group. In a column group, each process is in charge of a sub-domain, where the grid points are further grouped into grid blocks. A process traverses through its grid blocks using threads and in a grid block each thread executes SIMD instructions to update multiple grid points at the same time.

We have been developing efficient methods to speed up SpMV such as SFCs based grid partitioning and the use of OpenMP task-based parallelism. With the current capability of PARSEC, we have solved the electronic structure of systems of silicon nanocrystals with up to 26,000 atoms (or roughly 100,000 electrons) (Figure 2). We were able to observe the evolution of the density of states of silicon nanocrystals to the bulk limit. This size of system is by no means near a maximum with current computational resources. Systems with over several hundred thousand electrons have been run.

Following the inexorable trend, future computing power will be greatly enhanced; however, the communication speed between computing units and memory will still be limited. As a result, minimizing data transfer will remain a focal point. A worthy area of investigation will target the possibility of duplicating variables to decouple computation and communication and maximize their overlap.

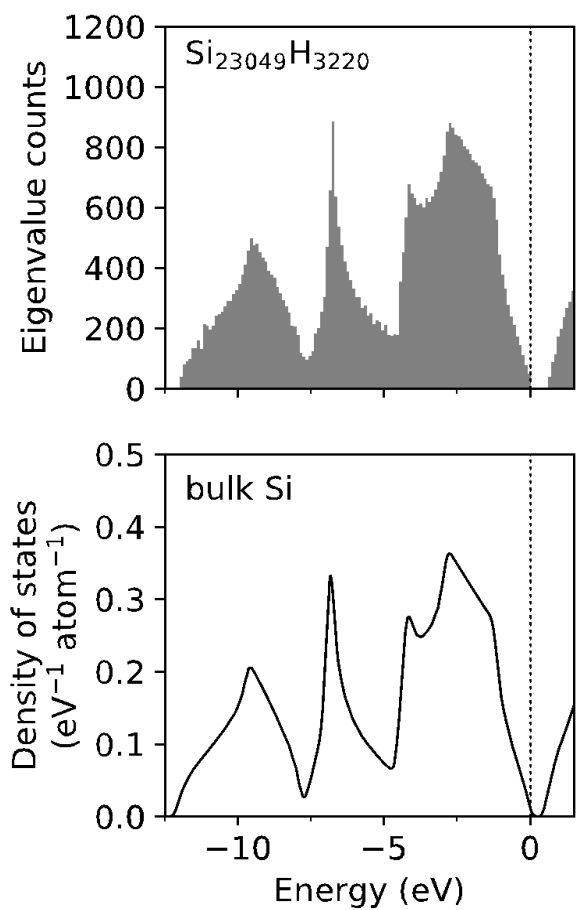

**Figure 2.** Density of states of a large silicon nanocrystal and the bulk silicon. The energy of the highest occupied state is set to 0 eV. The histogram bin width and the standard deviation of the Gaussian functions for convolution are 0.1 eV. Reproduced from Ref. [7] with permission from the American Chemical Society.

**Concluding Remarks**

Real-space formalisms have advantages in simplicity and ease of implementation. Moreover, owing to scalability, they have been employed for some of the large systems explored to date. The combination of real-space DFT and pseudopotentials constitutes a powerful and elegant tool for discovering and understanding new materials often on par with experiment.

PARSEC uses a high-order finite-difference method to discretize the physical space. The code supports a variety of restricted dimensionality for nanoscale structures, including spin-polarized calculations for magnetic materials and systems with neutral and charged defects. For the eigensolver, CheFSI is an efficient algorithm and a good match with real-space methods implemented on highly parallel platforms.

In the next generation code, we plan to support high-level exchange-correlation functionals to provide an option to include more accurate description for correlated physics problems. We also plan to support additional pseudopotential types and formats to facilitate a wider range of user needs. We will couple these thrusts to better computations of high-order forces and mixers, SFCs based grid partitioning for non-orthogonal lattices, and the use of GPUs.

To meet the challenges of exascale computing, we plan to expand the current parallelization scheme—from the topmost level of the spectrum of the Hamiltonian, Kohn–Sham states, down to grid blocks and grid points. With this multilevel parallel model, we hope to capitalize the opportunities provided by various hardware upgrades.

We are looking forward to working with the ever-growing electronic structure community to make use of the opportunities of the coming exascale era.

**Acknowledgements**
*We acknowledge support from a subaward from the Center for Computational Study of Excited-State Phenomena in Energy Materials (C2SEPEM) at LBNL, which is funded by the U.S. DOE, Office of Science, Basic Energy Sciences, Materials Sciences and Engineering Division under contract no. DE-AC02- 05CH11231, as part of the Computational Materials Sciences Program. Computational resources are provided by the National Energy Research Scientific Computing Center (NERSC) as well as Texas Advanced Computing Center (TACC). MD acknowledges support from the "Characteristic Science Applications for the Leadership Class Computing Facility" project, which is supported by National Science Foundation award #2139536.*

## 10 – The Qbox first-principles molecular dynamics code
François Gygi, University of California, Davis

**Background and Current Status**
First-principles molecular dynamics (FPMD) simulations are an essential tool for computational modelling of complex materials. Qbox is a C++/MPI/OpenMP implementation of first-principles molecular dynamics based on the use of pseudopotentials and the plane wave basis set. It was designed [1] for scalability on thousands of tasks, involving tens of thousands of processor cores. Qbox is routinely used to perform molecular dynamics (MD) simulations of systems including several hundred atoms. Its features include constant-temperature (NVT) and constant-pressure (NpT) MD simulations, the computation of maximally localized Wannier functions (MLWF), hybrid-DFT exchange-correlation functionals, and the computation of electronic response to arbitrary periodic perturbations. Qbox also implements the recursive subspace bisection (RSB) algorithm that allows for an efficient computation of the exchange energy in hybrid-DFT simulations with a controlled accuracy [2]. A notable Qbox feature is a client-server interface which allows for its use as a "DFT engine" driven by another program. This interface has enabled coupling to other software for efficient sampling of free energy surfaces, path-integral molecular dynamics (PIMD) simulations, and the computation of excitation energies using the Bethe-Salpeter equations. This approach relies on the development of flexible interoperable software components rather than integration of all features into a single code [3].

Sampling of free energy surfaces is often necessary in the study of systems including hundreds of atoms, that typically exhibit a complex energy landscape. The presence of energy barriers makes this exploration inefficient using plain MD. Advanced sampling methods become necessary to obtain an accurate picture of the free energy surface and compute e.g. reaction barrier heights. Qbox addresses this challenge by coupling to the Software Suite for Advanced Generalized Ensemble Simulations (SSAGES) [4] through its client-server interface. This coupled approach has been used to explore the free energy surface of a dipeptide [5], catalytic reactions on a metal surface [6], and the conformation of gold clusters [7]. In such coupled Qbox-SSAGES simulations, multiple instances of Qbox are "driven" by SSAGES in order to improve the efficiency of statistical sampling. In applications to PIMD simulations, Qbox was coupled to the i-PI software [8] that implements path integral sampling and various generalized Langevin thermostats. This was used to study nuclear quantum effects in diamond and amorphous carbon [9]. Finally, Qbox was coupled to the WEST code to compute electronic response integrals needed in a Bethe-Salpeter calculation of optical excitations [10]. This strategy of coupling Qbox with other interoperable software has considerably extended the range and accuracy of FPMD simulations [3].

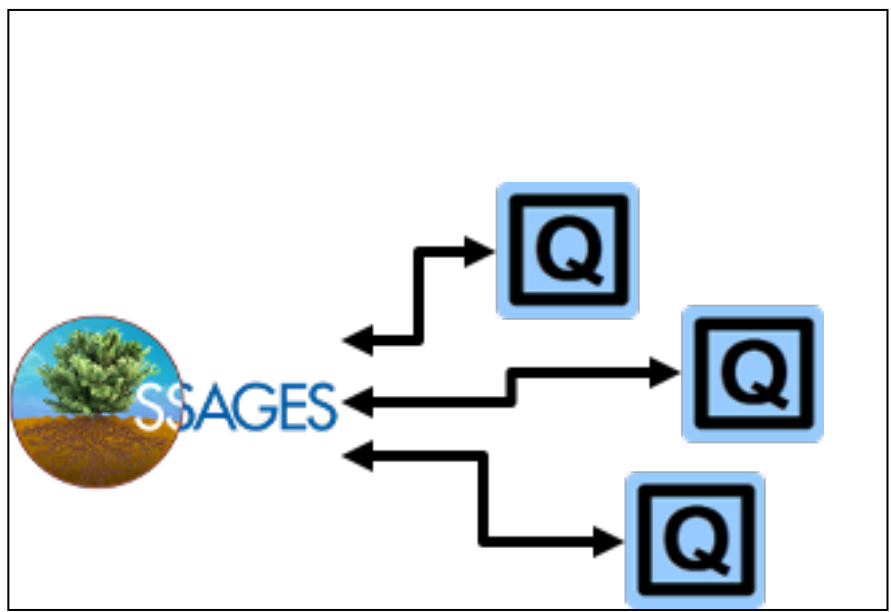

**Figure 1.** Schematic representation of a coupled Qbox-SSAGES simulation. The SSAGES program is driving multiple instances of Qbox processes. Each Qbox instance represents a separate copy of the system (or "walker") used in the exploration of the free energy surface. Atomic positions generated by SSAGES are sent to Qbox which returns energies and forces computed within DFT.

**Development Priorities**

The simulation of systems of increasing size usually implies the need for longer simulations in order to reach equilibrium conditions. Together with the inherent high cost of FPMD simulations, this puts a high premium on performance optimization. Reduction in the time needed to obtain the electronic ground state is a prime target of optimization. Qbox offers several choices of algorithms for the solution of the Kohn-Sham equations, the most commonly used being a preconditioned block Jacobi-Davidson method. Further performance enhancements are needed such as the efficient implementation of a parallel deflation algorithm and adaptive preconditioners. Furthermore, the charge density mixing algorithm used between SCF iterations in Qbox is based on a straightforward use of the Anderson acceleration algorithm (which was shown to be equivalent to the LBFGS algorithm). However, the parameters used in this approach, such as the dimension of the density subspace search, the parameters of Kerker mixing and the mixing coefficient are not optimal for all possible systems, and sometimes lead to convergence failure in large systems. Thus the exploration of robust and adaptive algorithms is an important development goal. Furthermore, wave function extrapolation algorithms can significantly impact performance during Born-Oppenheimer MD simulations by providing an accurate starting point for a subsequent Kohn-Sham SCF computation. The current extrapolation algorithm used in Qbox is a simple linear extrapolation preceded by subspace alignment. While some extensions to more complex extrapolation schemes have been proposed by some authors, a consensus has not been reached on the optimal way to extrapolate wave functions, and further exploration of these algorithms is warranted.

In addition to the optimization goals mentioned above, more immediate attention will be paid to the optimization of on-the-fly computation of polarization and polarizabilities during MD simulations, which directly affects the performance of simulations of the infrared and Raman spectra. The current Qbox implementation relies on a finite difference approach for the computation of polarizability, which can likely be accelerated using initial estimates from perturbation theory. Last but not least, the general question of the potential use of mixed floating-point precision algorithms in the context of Kohn-Sham solvers must be explored, in particular in view of the increasing imbalance of the cost of communication and computation on modern architectures.

**Meeting the Exascale Challenges**

Future exascale computers (with the notable exception of Fugaku installed at RIKEN in Kobe, Japan) are expected to achieve their peak performance using Graphical Processing Units (GPUs). This architectural development continues the trend of increasing aggregate peak floating point performance while the bisection bandwidth and byte/flop ratio continue to decrease. This poses a particular challenge for the implementation of algorithms that are not naturally embarrassingly parallel. The parallel solution of the Kohn-Sham equations in the plane wave basis is a tightly coupled problem involving frequent communication between different tasks, notably during parallel fast Fourier transforms (FFT). It also includes several computations that can be expressed as matrix products, an operation that has been traditionally thoroughly optimized and achieves near-peak performance.

On a traditional CPU-based platform, Qbox minimizes communication by partitioning processors into groups that host independent subtasks. For example, the solution of the Kohn-Sham equations for different k-points in the Brillouin zone, or for different spin indices, are largely independent. Furthermore, the computation of the electronic charge density requires Fourier transforms of each electronic orbital, leading to hundreds of independent tasks. Qbox currently divides tasks into a four-dimensional process grid in which plane wave coefficients, band, k-point and spin indices are distributed to different processors. FFT operations only involve communication along the first dimension of the process grid. Similar strategies are used to minimize communications in other parts of the computation.

On a mixed CPU-GPU architecture, communication between CPU and GPU, and between GPUs occurs at a much reduced rate compared to the memory bandwidth within a GPU. This makes the implementation of a plane wave Kohn-Sham solver difficult. A prototype of Qbox was developed for operation on NVIDIA A100 GPUs, demonstrating that the performance of a single GPU can be successfully exploited using hand-coded kernels and vendor-supplied libraries for FFT and matrix operations. The extension to multiple GPUs remains a challenge and will likely require substantial redesign of the parallelization strategy.

The constant evolution of GPU architectures, and the concomitant need for redesign implies that such an ad-hoc approach to specialize a code for a particular GPU model is unsustainable. As multiple vendors currently offer multiple GPU architectures (NVIDIA, AMD, Intel) that all rely on different programming models, the task of porting and maintaining a code on all such architectures appears daunting. Ideally, the adoption of a single, directive-based, portable programming model such as OpenMP, OpenACC or OneAPI should make it possible to maintain a single branch of a code. This approach must however be tested to verify that the resulting performance is comparable to hand-written specialized versions of the code. As this evaluation requires considerable rewriting, it has not yet been performed on Qbox. We plan to explore an alternative way to facilitate portability. Using the polymorphism and inheritance features available in C++, the architecture-dependent, hand-optimized parts of the code may be limited to specific platform-dependent derived classes, which are selectively instantiated at run time according to the available hardware. This approach can in principle preserve the performance of hand-optimized code while limiting the amount of such specialized code.

As we pursue the development of GPU-enabled algorithms, a more direct use of exascale platforms can be made in the context of FPMD simulations coupled with advanced sampling. Using multiple replicas, or walkers, in sampling algorithms leads to nearly independent problems that can easily take advantage of large computing resources. It is therefore likely that future simulations will involve such weakly coupled FPMD problems and will allow for improved accuracy and reduced statistical error.

**Concluding Remarks**

The developers of first-principles simulation codes face a growing demand for numerous new additional features, while having to maintain a high-performance implementation. The human resources needed to manage such software development invariably grow with its complexity, which increases the maintenance cost over time. We have explored an approach involving multiple interoperable software components that helped limit the cost of software development, while allowing for extension of the available features. This approach also led to an efficient use of the latest developments implemented by the various software teams involved, e.g. the inclusion of the latest sampling algorithms in SSAGES, or the latest thermostats implemented in i-PI. Finally, the evolution of first-principles simulation codes has been strongly affected by changes in hardware architecture. The current multiplication of hardware architectures puts the emphasis on the design of a flexible software architecture that avoids redesign or confines it to limited parts of the code. History has shown that simulation codes must adapt to multiple changes in hardware architecture during their lifetime. We expect that the design of a flexible software architecture will become the most important feature of a simulation code as further changes in architecture will no doubt appear at a sustained pace.

**Code availability**

The examples mentioned in this work use open-source codes: Qbox (http://qboxcode.org ), WEST (http://west-code.org ), SSAGES (https://ssagesproject.github.io/), i-PI (http://ipi-code.org ).

**Acknowledgements**

This work is supported by the Midwest Center for Computational Materials (MICCoM) as part of the Computational Materials Sciences Program funded by the U.S. Department of Energy, Office of Science, Basic Energy Sciences, Materials Sciences, and Engineering Division through Argonne National Laboratory, under Contract No. DE-AC02-06CH11357.

## 11 – Q-Chem: High-Efficiency Software for Quantum Molecular Workflows

John M. Herbert
Department of Chemistry & Biochemistry, The Ohio State University, Columbus, OH 43210 USA

**Background and Current Status**

Unlike much of the other software described in this article, Q-Chem is first and foremost a *molecular* quantum chemistry code, designed to describe a finite system using atom-centered Gaussian basis functions and featuring both density functional theory (DFT) as well as correlated, post-Hartree–Fock wave function models. Q-Chem is also a relatively mature code, having been under continuous development since 1993, albeit with an evolving code base and an infrastructure representative of modern programming practices [1]. One of Q-Chem's biggest strengths is its large academic developer base, evidenced by a list of more than 200 co-authors for the latest release [1]. This provides for tremendous diversity of features and methods, some of which are not widely available in other codes.

From a user perspective, Q-Chem's target audience has historically been chemists and other molecular scientists interested in the structure, reactivity, and spectroscopy of single molecules or at most small clusters of molecules. For larger systems, Q-Chem can function as the quantum engine of a quantum/classical (QM/MM) approach [2,3], or else a QM/QM embedding approach [1]. That emphasis has steered Q-Chem's design philosophy, which has long targeted performance on workstation-type hardware, with particular focus on single-node, shared-memory parallelism. Its engine for computing electron repulsion integrals over Gaussian basis functions is highly optimized to minimize both memory operations and floating-point operations, via a meta-algorithm that determines the optimal approach at runtime based on the characteristics of the requested basis set [4]. An analogous meta-algorithm for graphics processing units (GPUs) [5], which can be interfaced with Q-Chem [5], also determines whether single- or double-precision arithmetic should be used for each class of integrals. The result is near-optimal single-processor efficiency, with multithreaded parallelization accomplished via the OpenMP protocol.

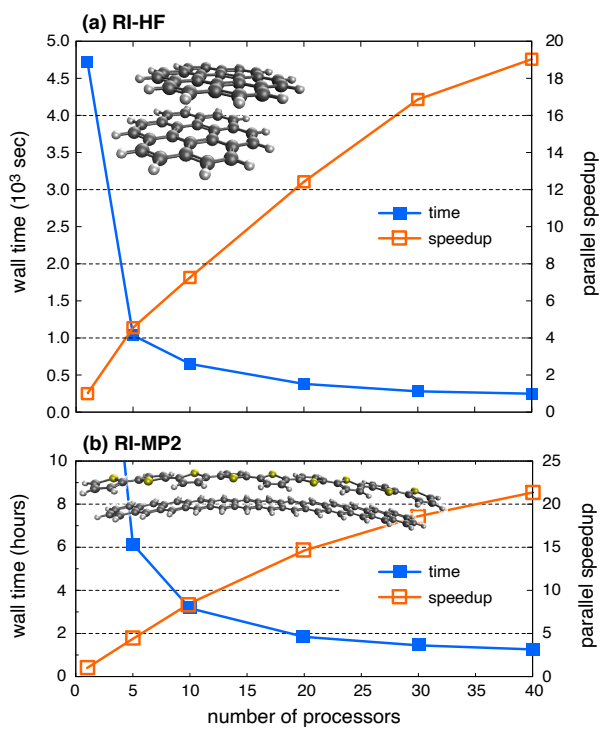

**Figure 1.** Strong scaling data for (a) Hartree-Fock calculations of coronene dimer using the def2-TZVPD basis set (72 atoms and 1,992 basis functions) and (b) MP2 calculations of a graphene/oligothiophene dimer using the same basis set (149 atoms and 4,194 basis functions). Both calculations use resolution-of-identity (RI) integrals, with a screening threshold of $10^{-12}$ and a SCF convergence threshold of $10^{-5}$ Ha. Parallel speedup data are shown on the right, relative to single-processor timings.

**Development Priorities**

Single-node performance of Q-Chem for single-point energy calculations is demonstrated in Fig. 1, at the level of second-order Møller-Plesset perturbation theory (MP2). The resolution-of-identity (RI) approximation is used for the electron repulsion integrals in both the Hartree-Fock (HF) and MP2 calculations. Timing data are shown for a tight integral screening threshold of $10^{-12}$ as numerical linear dependencies can significantly hamper SCF convergence in systems of this size, if thresholds are set too loose. As exascale platforms turn applications of this size (or larger) into routine affairs, one may anticipate that drop tolerances will need to tighten relative to that values of $10^{-8}$–$10^{-9}$ a.u. that are sometimes used at present, which work acceptably well for medium-size molecules.

Moving forward, Q-Chem plans to pursue a distributed-memory, hybrid OpenMP/MPI parallelization strategy that will facilitate applications to systems that are simply too large for the single-node approach. This will be needed for periodic DFT calculations, a functionality that has recently been added [6]. The parallelization strategy to be pursued is a conservative one, however, designed to avoid sacrificing Q-Chem's excellent single-node performance. This will admittedly limit the scalability but will preserve Q-Chem's outstanding price-to-performance ratio that makes large systems accessible *without* the need for supercomputer or leadership-class computing resources.

Q-Chem's design philosophy is well-suited for high-level calculations of spectroscopic properties of molecules, which might themselves be embedded in a larger framework. (In addition to conventional QM/MM frameworks, projector-based wave function-in-DFT embedding methods are also available, along with some types of frozen-density embedding [1].) Computational spectroscopy has long been an area of strength for Q-Chem, as a variety of correlated excited-state wave function models are available along with several different DFT-based approaches including both real-time and linear-response time-dependent DFT, as well as ΔSCF and transition-potential methods [1]. Given Q-Chem's design philosophy as well as the structure of the existing code, it is unrealistic to imagine that Q-Chem will soon become a code that scales to thousands of processors. Instead, its power lies in leveraging excellent single-node performance to tackle large problems by breaking them up into smaller ones, or in other words, parallelization at the level of workflows. Fragment-based approaches to quantum chemistry [7], which seek to use physics-based approximations to turn large problems into collections of much smaller ones, represent the best strategy to use a code like Q-Chem to attack large problems.

**Meeting the Exascale Challenges**

In this author's opinion, efforts to demonstrate that electronic structure software can run on thousands or hundreds of thousands of processors often feel performative, with funding agencies or program officers as the presumed audience. Typical users do not easily have access to this kind of computing resources on a routine basis. Furthermore, what is often overlooked in brute-force scaling demonstrations of this sort is whether resources are being used *efficiently* to solve the problem at hand. In this context, "resources" really means the electricity required to operate the hardware and from that point of view, the cost of a given calculation would best be measured not by wall-clock time but instead by power consumption or carbon footprint. That metric is tricky to evaluate on shared computing resources but total CPU time (aggregated across all processors) is readily available and can serve as a stand-in that reflects the true cost of a given calculation. Wall-clock time is a more selfish metric that reflects only a single user's time-to-solution, and wastes resources if parallel efficiency is low.

With the aim of using fragment-based methods to target large systems [7], Fig. 2 shows some timing data for Hartree-Fock calculations on full proteins. Alongside the conventional results are data from a fragment-based approach, the pair–pair generalized many-body expansion or pp-GMBE(2) [8]. The latter approach does not require any single calculation that is larger than four amino acids yet provides relative conformational energy profiles that are faithful to the full macromolecular result [7,8]. By the nature of the fragmentation approximation, wall-clock time can be reduced to the cost of a single tetrapeptide calculation if sufficient hardware is available. Nevertheless, the aggregate CPU cost is considerably lower for the standard Hartree-Fock calculation than for its fragment-based approximation, even for proteins with well over 1,000 atoms!

Fragment-based approximations have grown in popularity in recent years but many of the proposed approaches fail to maintain good fidelity with respect to the supersystem calculation they aim to approximate [7,9]. Conversely, methods that are faithful to the supersystem result (to within ~1 kcal/mol accuracy, say) have proven to be more expensive when cost is measured in total aggregate CPU time [7–9], as seen for proteins in Fig. 2. Recently, however, significant progress has been made in reducing the computational expense of the fragment-based approaches, while preserving high fidelity, via energy-based screening of the fragments at a low level of theory [7,10]. The result is a method that maintains ~1 kcal/mol accuracy for challenging problems (such as relative energy differences for proton ordering in water clusters), yet are more affordable than the full system calculation starting at relatively small systems [10].

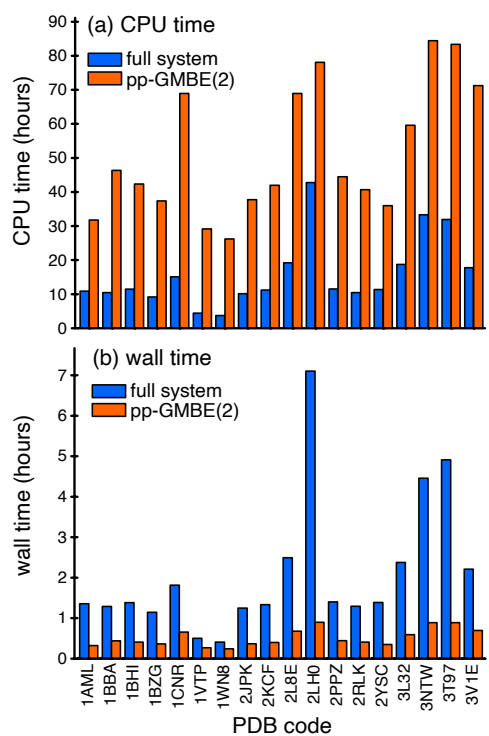

**Figure 2.** (a) Total CPU time and (b) wall-clock time for Hartree-Fock/6-31G* calculations on various proteins, comparing a full system calculation to a fragment-based calculation [pp-GMBE(2)] that preserves relative energies at the level of 1–3 kcal/mol while requiring no single electronic structure calculation larger than four amino acid residues. Full-system calculations were performed on a single 12-core compute node whereas pp-GMBE(2) calculations used 10 of the same nodes (120 cores). Adapted from Ref. [9]; copyright 2016 American Chemical Society.

**Concluding Remarks**

At first glance, Q-Chem might seem ill-suited for the "exascale era". However, multi-node parallelization at the level of workflows (rather than individual energy or gradient evaluations) can be an effective and efficient route to supercomputing in quantum chemistry, especially if the cost of a given calculation is measured by its carbon footprint, or in other words, by total CPU time aggregated across all processors. This is no less true in an era of machine learning and other "big data" approaches to computational science, which place a premium on efficient generation of large data sets. Even in the exascale era, the author predicts that a significant amount of computing will continue to be done on workstations, for which Q-Chem is highly optimized.

**Acknowledgements**

Work by the author on fragment-based quantum chemistry has been supported by the U.S. Department of Energy, Office of Science, Office of Basic Energy Sciences, Division of Chemical Sciences, Geosciences, and Biosciences, under Award DE-SC0008550. The author serves on the board of directors of Q-Chem Inc.

## 12 – Quantum Espresso

S. Baroni, SISSA-ISAS and IOM-CNR, Trieste, Italy
P. Delugas, SISSA-ISAS, Trieste, Italy
P. Giannozzi, University of Udine and IOM-CNR Trieste, Italy

**Background and Current Status**

Quantum ESPRESSO (QE) is a distribution – an integrated suite - of codes for electronic-structure calculations of materials properties, based on density-functional theory (DFT), pseudopotentials, and plane waves [1]. The roots of QE are in solid-state physics: the oldest parts of QE have been in use and under development since the mid-80's, originally applied to compute structural and electronic properties of simple semiconductors. In particular, the linear-response and molecular dynamics codes in QE derive from the original implementations of density-functional perturbation theory [2] and of Car-Parrinello molecular dynamics [3], respectively. QE has since been extended to cover a much wider class of materials and properties, providing basic functionalities—structural optimization and first-principle molecular dynamics with semi-local, nonlocal, Hubbard-corrected, and hybrid functionals—as well as more advanced ones—e.g.: nudged elastic band, linear response—for materials science, geophysics, chemical and biological physics, and many branches of engineering [4]. The current version (7.1) allows the computation of the vibrational spectra (phonons), of electron-phonon interaction coefficients via the EPW package [5], of electronic excitations with time-dependent DFT, and of spin-wave excitations (magnons) [6].

The open character of the development and the considerable work done to enable various forms of interoperability with external software make QE suitable both as a building block for more complex software suites and as a generator of the starting electronic structure for advanced theories like many-body perturbation theory or quantum Monte Carlo.

QE is developed following modern software best practices, including continuous integration and extensive testing. It has a significant number of active developers and a rather large user basis, ensuring a very careful monitoring of the correctness of the results. QE is written in modern Fortran (up to 2008 standard) but its coding reflects as closely as possible the underlying physics, thus allowing to make experiments, customizations, extensions, new developments, without a too high learning barrier.

Efficiency on available computers has always been a major concern for developers. QE has been working on the entire range of computer hardware available at any given time: from personal workstations and laptops to earlier vector supercomputers and parallel machines, up to the most recent hybrid accelerated architectures. Excellent performance on various kinds of pre-exascale machines with Nvidia GPUs has already been achieved [7], and rapid progress is being done for other kinds of accelerated hardware.

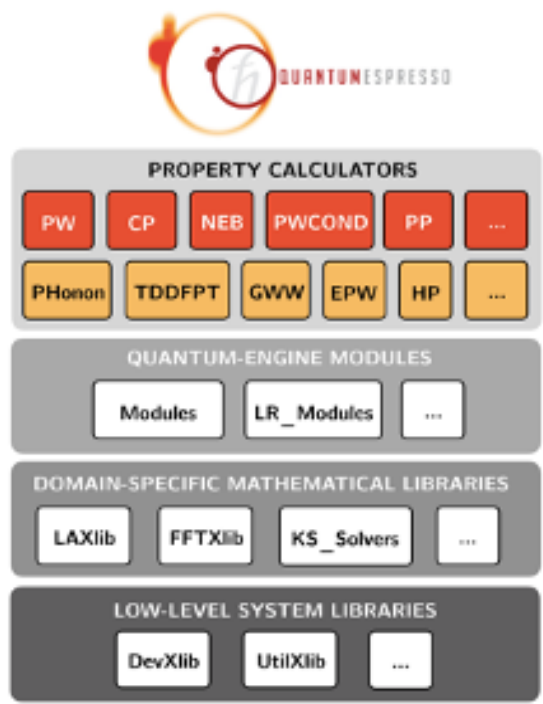

**Figure 1.** Schematic illustration of the structure of the QUANTUM ESPRESSO distribution. Reprinted with permission from Ref.[7].

**Development Priorities**

One of the original goals of QE was to become a hotbed for innovation. While it is hard to predict which new developments will be proposed by the scientific community in the future, one can be confident that a primary goal of QE will always be to make implementation of new algorithms, methods, and theories as easy as possible. In order to achieve such a goal while keeping the complexity of QE under control, an open and sustainable development model was devised. An increasing number of QE components are being encapsulated into reusable libraries and Fortran modules, thus making new developments easier to implement (Fig. 1).

Among development priorities for the next few years, we mention in particular the implementation of new advanced functionals, notably Hubbard-corrected, Koopmans-compliant, meta-GGA and nonlocal van-der-Waals functionals. Moreover, for several already implemented cases of advanced functionals, improvements to their numerical stability and speed are needed in order to make such advances more usable and useful. For hybrid functionals, in particular, it is planned to use more extensively the localization of wavefunctions to boost the performance. [8]

For non-collinear magnetic or even conventional (e.g., GGA) calculations, improving the robustness of self-consistency for both ground-state and linear-response calculations is one of the main goals for the near future. This is especially important given the increasing usage of QE for high-throughput calculations and for machine-learning techniques, requiring large quantities of reliable data to be produced and reproducible in an automated way. A global minimization approach already exists in the QE distribution, but is limited to the Car-Parrinello code. It will be ported, possibly encapsulating it in a module, to the main electronic-structure code, as a more robust alternative to the traditional, and usually faster, self-consistent procedure.

In the field of spectroscopic properties, a major development goal is molecular dynamics on the excited state, using time-dependent DFT. This is a very important tool for the understanding e.g. of photochemical reactions and photovoltaic processes.

A further field of active development is the introduction of new "multi-scale" methods, or the consolidation and extension of existing ones. With such expression we mean methods in which the effect of the environment surrounding the quantum-mechanical system is introduced in some approximate way. In addition to the prototypical QM-MM molecular dynamics, we mention here SCCS (Environ) [9], ESM and 3D-RISM [10].

A development priority of more technical character, but one related to some of the mentioned applicative objectives, is making QE more interoperable with other languages (notably modern object-oriented ones like python) and with other pieces of software. The effort in this field is ongoing since several years and has already produced significant results, such as structured (xml, hdf5) I/O files and various degrees of actual interoperability with other software. As a further step forward in this direction, a set of documented API's for calling QE subroutines will be introduced, making QE routines more easily accessible from other codes or directly from Python-based software packages.

**Meeting the Exascale Challenges**

For conventional machines based on many multi-core CPU nodes, the main parallelization strategy implemented in QE – dividing plane-wave components, in both real and reciprocal space, across processors - is well established and effective. Strong scaling is limited by the size of the real- and reciprocal-space grids and by the parallel distributed Fourier Transforms. QE uses both MPI and OpenMP parallelization and introduces several additional parallelization levels to achieve better scaling. In practice, the basic self-consistency or molecular-dynamics calculations scale well up to dozens of processors for small-medium size system (described by supercells containing hundreds of atoms), up to a few thousand processors for large-scale calculations (supercells up to a few thousand atoms). Scaling beyond such limits strongly depends upon the specific calculation and in particular upon the availability of additional parallelization levels (see Fig.2).

The push towards hybrid and accelerated processors (e.g., GPGPU) of the last few years has been addressed in QE with a variety of approaches. For Nvidia GPUs, the first approach was a porting based on the CUDA Fortran extension [7]. The result was very gratifying in terms of performance, much less so in terms of portability to other architectures and also of maintainability (due to extensive code duplications, imposed by limitations of CUDA Fortran).

More recently, most NVidia-specific code has been moved to OpenACC, with no loss of performance and much reduced code duplication. An ongoing effort is under way, and almost completed, to extend the NVidia porting with OpenACC to the entire suite and not only to the most used basic components. Work is also under way to implement OpenMP for other hybrid architectures, in the hope that the latter will become the open standard. The pathway for further porting becomes much easier and is limited only by the availability of compilers enabling OpenACC and OpenMP.

The porting on accelerated architectures is especially important in view of the future availability of "exascale" machines, that will be presumably based on such architectures. The prospect of a machine that is capable of $10^{18}$ operations per second, and of the results that could be obtained with it, is exciting. Translating such unprecedented computer power into actual scientific results is however a challenge, and not just for code developers. One has first to identify which kind of calculations can actually profit from exascale capabilities. The vast majority of basic DFT simulations— structural optimizations, first-principle molecular dynamics calculations—may neither exploit nor really need such a huge computing power. Improving the performances and usability of additional parallelization levels, in particular over Kohn-Sham states, will be key for better exploiting future architectures and a top development priority of QE. In practice, this means a systematic distribution over nodes of all arrays, a careful optimization of inter-node communications, the removal of unneeded synchronization points and overlapping computation and communications whenever possible.

Scientifically relevant cases requiring many loosely coupled calculations of different systems, or different replicas of the same systems, are often encountered. Typical cases include high-throughput calculations (many independent configurations), calculations of phonon spectra (many irreproducible representations and wave-vectors) and nudged elastic band (several images on a string). Those cases may potentially require exascale capabilities. QE can already address this kind of massive parallelization by exploiting additional levels of parallelism.

Among single calculations potentially requiring exascale capabilities within the scope of QE, we mention calculations with hybrid functionals, where a wise usage of parallelization over Kohn-Sham states may allow to perform highly accurate calculations on large unit cells, currently not feasible, using a very large number of processors [8].

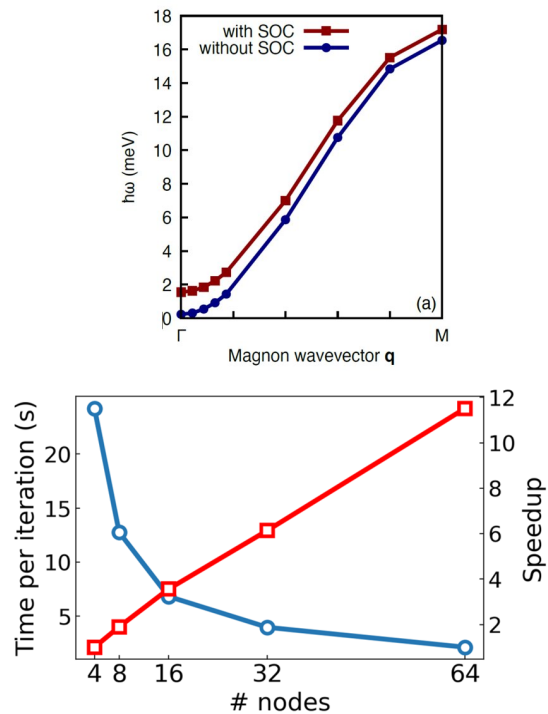

**Figure 2.** Top: magnon dispersions in CrI$_3$ (reprinted with permission from Ref.[6]). Bottom: scaling performances of the TURBO-MAGNON code for each magnon wavevector. Each node has two 24-core CPUs and executes 24 MPI processes and 2 OpenMP threads.

## Concluding Remarks

The original motivation to develop QE was to provide a unified set of software tools for a small but very active community of experts in first-principle methodologies and simulations. The scope of QE has grown during the years and its ambition is now to cater to a wider community of scientists and engineers, working in different fields, that includes non-specialists as well. For this reason the development of new capabilities is accompanied by an intense dissemination activity, especially targeting younger scientists and aimed towards spreading knowledge and expertise in first-principle DFT calculations and in their usefulness. We consider this activity an integral part of the "QE experience" and one of the most successful aspects of QE.

## Acknowledgements

This work was partially funded by the EU through the MaX Centre of Excellence for HPC applications (Project No. 824143). Financial and administrative support from the Quantum ESPRESSO Foundation is also gratefully acknowledged.

## 13 – SPARC: Advanced algorithms for systems large and small

*Phanish Suryanarayana[1], Qimen Xu[1], and John E Pask[2]*

[1]Georgia Institute of Technology  [2]Lawrence Livermore National Laboratory

**Background and Current Status**
SPARC [1, 2] (https://github.com/SPARC-X/SPARC) is an open-source, real-space density functional theory (DFT) code that accommodates Dirichlet and Bloch-periodic boundary conditions, enabling the treatment of finite, semi-infinite, and charged systems, as well as bulk 3D systems. SPARC employs the finite difference method, wherein quantities of interest are discretized on a uniform real-space grid and convergence is controlled by a single parameter. The finite-difference method's simplicity, locality, and freedom from communication-intensive transforms enables efficient implementation on large-scale parallel computers. Because the representation is maximally local in real space, modern $O(N)$-scaling as well as traditional $O(N^3)$-scaling methods are readily implemented.

Current features of SPARC include:
- Applicable to isolated systems such as molecules as well as extended systems such as crystals, surfaces, and wires.
- Local, semilocal, and nonlocal (including hybrid) exchange-correlation functionals.
- Standard ONCV pseudopotentials, including nonlinear core corrections.
- Calculation of ground state energy, atomic forces, and stress tensor.
- Structural relaxation and ab initio molecular dynamics (NVE, NVT, and NPT).
- Spin polarized and unpolarized calculations.
- Spin-orbit coupling.
- Dispersion interactions through DFT-D3, vdW-DF1, and vdW-DF2.
- Linear-scaling Spectral Quadrature (SQ) method [3, 4].
- Discrete Discontinuous Basis Projection (DDBP) method [5].
- Symmetry adaption for cyclic and helical symmetries [6].
- Orbital-free DFT with TFW, WT, and WGC kinetic energy functionals.
- MATLAB version available for rapid prototyping: M-SPARC [7].

SPARC is portable and straightforward to install, use, and modify, with external dependencies limited to industry standard BLAS, LAPACK/ScaLAPACK, and MPI. It has been extensively validated and benchmarked against established planewave codes [1, 2], where SPARC has shown to be an order of magnitude faster, with increasing advantages as the number of processors is increased [1]. It can efficiently utilize modest as well as substantial computational resources, with parallel scaling bringing solution times to about a minute for systems with O(500-1000) atoms (Fig. 1), and a few seconds for O(100-500) atoms [1]. Using the O(N) SQ method, it has been scaled to system sizes of over a million atoms (Fig. 2). Unique features of SPARC enable the study of extreme conditions of temperature/pressure [8] as well as systems with cyclic/helical symmetry, as found in nanostructures intrinsically and subject to bending and torsional deformations [9].

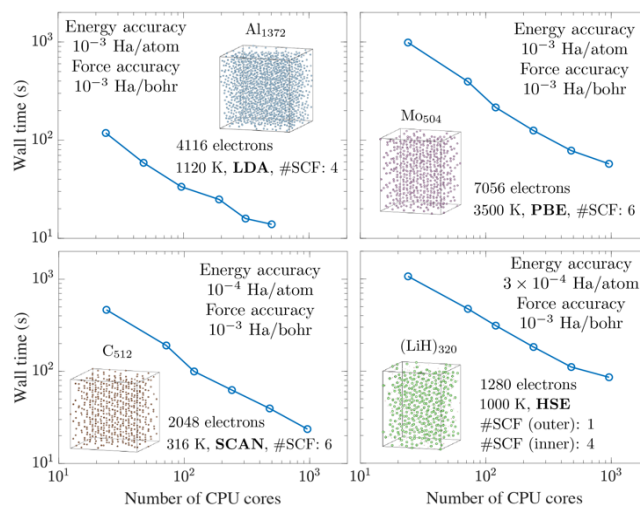

**Figure 1.** Efficiency and scaling of SPARC for an AIMD step with a range of exchange-correlation approximations, obtained on Phoenix computer at Georgia Tech.

**Development Priorities**

SPARC is under active development to increase length/time scales and level of theory accessible. Key developments targeted in the next 5 years include:

*O(N) DFT*: The main bottleneck in DFT calculations is the solution of the Kohn–Sham eigenproblem, which not only scales as $O(N^3)$ with system size but also involves global communications between processors, limiting parallel scalability. As a result, much research has been devoted to the development of O(N) methods, with reduced scaling achieved by exploiting the locality of electronic interactions in real space [10]. While these efforts have yielded significant advances, concerns remain regarding the accuracy, stability, and large prefactors, particularly for metallic systems. The O(N) SQ method implemented in SPARC addresses the first two concerns, however its prefactor increases with decreasing electronic temperature. To address this, the DDBP method will be employed to reduce the prefactor by orders of magnitude.

*Density functional perturbation theory (DFPT)*: DFPT is an elegant approach for determining the system's response to perturbations in electronic structure calculations, without the need for large supercells and/or a series of ground state calculations involving unperturbed and perturbed systems. DFPT has found a number of applications including structural stability, elastic moduli, flexoelectric coefficients, Raman spectra, electro-optic coupling, ferroelectric transitions, transport properties, and thermodynamic properties. Conventional DFPT calculations scale as $O(N^3-N^4)$, with a large prefactor associated with the solution of a linear system. $O(N^2-N^3)$ formulations of DFPT will be developed and implemented, with orders of magnitude reduction in prefactor achieved through the use of the DDBP method, and preconditioning techniques that exploit the similarity in linear systems solved.

*Random phase approximation (RPA)*: At the fifth and highest rung of Jacob's ladder of exchange-correlation functionals, RPA --- capturing Van der Waals interactions, free from self-interaction error, and applicable to small-gap and metallic systems --- is considered the gold standard for condensed matter systems in a number of research areas, particularly where there is demonstrated need to increase the accuracy of energies beyond traditional DFT. Conventional RPA calculations for the correlation energy scale as $O(N^4)$, with a large prefactor that grows as the fourth power of the number of grid points/atom. $O(N^2)$ formulations of RPA will be developed and implemented using the SQ approach in the DFPT framework, with orders of magnitude reduction in prefactor achieved through the DDBP method.

*On-the-fly machine learned force fields (MLFF)*: A variety of machine learning techniques have been developed to accelerate molecular dynamics simulations, leveraging the substantial data generated in the course of such simulations. This includes Gaussian process regression (GPR)-based on-the-fly MLFF, which has found a number of applications, including phase transitions and transport properties. However, this method is currently limited to system sizes that are of same order as typical DFT simulations, due to the cubic scaling bottleneck of GPR training, quadratic scaling of the feature vector with number of chemical elements, and the need for significantly more training configurations for systems with highly heterogeneous bonding. GPR-based on-the-fly MLFF schemes will be developed and implemented, with featurization schemes and hierarchical matrix algorithms that overcome the aforementioned bottlenecks.

**Meeting the Exascale Challenges**

The key computational kernel in real-space Kohn-Sham DFT calculations is the solution a large sparse nonlinear eigenproblem for the orbitals and corresponding eigenvalues. The number of orbitals and eigenvalues that need to be computed is proportional to the number of atoms/electrons in the system. Since these orbitals need to be orthogonal, the computational complexity of the kernel scales as $O(N^3)$ with number of atoms/electrons. This orthogonality constraint also limits parallel scalability, due to the need for global communication in parallel computations. Therefore, even with the advent of highly scalable eigensolvers, the efficient use of petascale and exascale machines presents a significant challenge.

However, the Kohn-Sham problem can be formulated in terms of the density matrix rather than orbitals and eigenvalues, from which quantities such as electron density, energy, atomic forces, and stresses can be determined directly. By exploiting the decay of the density matrix in real space, i.e., locality of electronic interactions, O(N) scaling can be achieved [10]. However, while significant progress has been made in the development of O(N) methods over the past two decades, petascale and exascale machines present new challenges. In particular, even though O(N) methods require minimal global communications, efficient large-scale parallelization poses a significant challenge due to complex communications patterns and load balancing issues associated with underlying localized orbital representations. In addition, the dramatically larger prefactors associated with present methods, for metallic systems in particular, significantly limits practical utility.

The SQ method in SPARC, which is applicable to metals and insulators alike, addresses the challenge for O(N) scaling and large-scale parallelization (Fig. 2). It is well suited to scalable high-performance parallel computing, with nearly all communications localized to nearby processors, with a pattern that remains fixed throughout the simulation. In particular, once the localized communication is complete, the calculations associated with each grid point are completely independent, whereby the SQ method naturally scales to the number of processors equaling the number of grid points, and beyond when an additional level of parallelization for computations at each grid point is implemented. Given that typical real-space DFT calculations employ O(500–30,000) grid points/atom, the SQ method is well suited to scale efficiently on petascale and exascale machines.

To reduce the prefactor of the SQ method, while retaining its parallel scalability and systematic convergence to $O(N^3)$ results, it will be implemented within the framework of the DDBP method. In particular, the DDBP method systematically reduces the dimension of the discrete real-space eigenproblem that must be solved by 1–3 orders of magnitude, which translates to a similar reduction in prefactor. Indeed, the generation of the DDBP basis — strictly localized, orthonormal, and discontinuous — scales as $O(N)$ with natural and efficient parallelism, given the multiple levels of parallelization available with all communications localized to nearby processors.

The implementation of SPARC on heterogeneous architectures, currently in progress, uses OpenMP 5.x features for accelerator devices, with minimal device-specific features. Leveraging the locality and multiple levels of parallelism available in the SQ and DDBP methods, working sets and data can be GPU-resident and efficient multi-GPU operation can be targeted while minimizing inter-node communication.

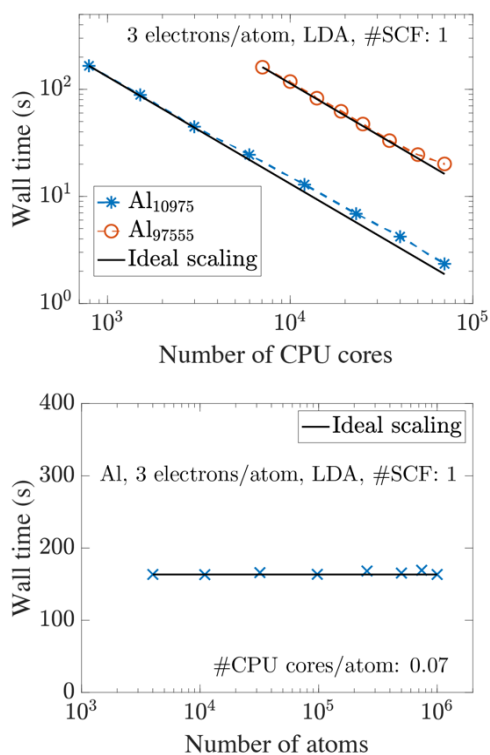

**Figure 2.** Strong and weak scaling of the O(N) SQ method in SPARC for a metallic system on Quartz computer at LLNL, with energies and forces computed to chemical accuracy. The largest system in the weak scaling contains 1,000,187 atoms.

**Concluding Remarks**

SPARC is an accurate, efficient, and scalable open-source electronic structure code, with many advanced features, able to efficiently leverage moderate and large-scale computational resources alike. It is straightforward to install, use, and modify, with minimal external library dependencies. It has shown to be an order of magnitude faster than state-of-the-art planewave codes, with a range of exchange-correlation functionals, and with increasing advantages as the number of processors is increased. In particular, SPARC efficiently scales to thousands of processors in regular operation, bringing solution times down to about a minute for systems with O(500-1000) atoms, and a few seconds for O(100-500) atoms. Using the O(N) SQ method, it has been scaled to system sizes of over a million atoms. Further reductions in solution times are on the horizon with the release of the DDBP method within SPARC. Given the excellent parallel scalability of SPARC, cutting-edge methods such as O(N) SQ and DDBP that are well suited to petascale and exascale machines, and variety of boundary conditions, SPARC promises to enable a number of new and exciting applications that were previously beyond reach.

**Acknowledgements**
This work was supported by grant DE-SC0019410 funded by the U.S. Department of Energy, Office of Science. This work was performed under the auspices of the U.S. Department of Energy by Lawrence Livermore National Laboratory under Contract DE-AC52-07NA27344, with support from the LLNL ASC/PEM and Multiprogrammatic and Institutional Computing programs. The authors thank the other developers for their contributions to SPARC.

## 14 — Towards exascale calculations of excited state properties with the WEST code

Marco Govoni[1,2], Victor Wen-zhe Yu[1], Giulia Galli[2,1]
[1]Materials Science Division, Argonne National Laboratory
[2]Pritzker School of Molecular Engineering, University of Chicago

**Background and Current Status**

Computer simulations based on first-principles electronic structure methods are widely adopted to explain, complement, and guide experiments in materials science. Applications in areas of energy sustainability and quantum information science require atomic-scale understanding of the ground-state and excited-state properties of complex heterogeneous systems, including nanostructures, surfaces, interfaces, and defects in solids. Despite the success of density functional theory (DFT) in describing a wide range of ground-state properties, many implementations of DFT are known to be inaccurate to describe excited states and strongly correlated systems. Many-body perturbation theory (MBPT) has been developed into mainstream methods, for example GW and the Bethe-Salpeter equation (BSE), that reliably describe a variety of charged and neutral electronic excitations. In addition, MBPT may serve as the basis for calculations of strongly correlated electronic states using Green's function embedding approaches.

In general, calculations based on MBPT methods are computationally more demanding than those using DFT. Conventional implementations of GW and BSE, for instance, exhibit a computational complexity that scales as $O(N^4)$ and $O(N^6)$, respectively, where N is the number of electrons in the system, posing difficulties in carrying out MBPT calculations for complex, heterogeneous materials.

We have implemented MBPT in an open-source software package named WEST (Without Empty States, http://west-code.org) [1] that is interfaced with the Quantum ESPRESSO (https://quantum-espresso.org) and Qbox (http://qboxcode.org) plane-wave pseudopotential codes. The key functionalities of WEST are summarized in Fig. 1. WEST adopts distinctive algorithms to circumvent computational bottlenecks commonly encountered in large-scale MBPT calculations. Specifically, $G_0W_0$ [1], BSE [2], electron-phonon [3], and quantum embedding [4] calculations are implemented in WEST without performing any summation over empty electronic states to obtain dielectric matrices and Green's functions, thus avoiding a severe computational burden present in traditional plane-wave based MBPT implementations. WEST utilizes spectral decompositions of density-density response functions, and compact basis sets of dielectric eigenpotentials, thus eliminating the need to store and invert large dielectric matrices. The $G_0W_0$ implementation is carried out using full integration over the frequency domain without using generalized plasmon-pole models to approximate frequency-dependent dielectric response functions. The $G_0W_0$ implementation of the WEST code was verified by comparing calculations using pseudopotentials to all-electron reference data [5] and generalized to include spin-orbit coupling. The solution of the BSE is implemented using a finite field approach and uses the recursive bisection method to reduce the scaling of the calculation [2]. The calculation of electron-phonon self-energies is implemented for ground states obtained either with semi-local or hybrid functionals.

WEST has been used to study excited states for a variety of systems, including molecules, nanoparticles, two-dimensional materials, spin defects in solids, liquids, amorphous, and solid/liquid interfaces. These systems are often represented with super-cells with tens of thousands of electrons and their finite temperature description may require averaging results over several configurations extracted from molecular dynamics simulations, thus leading to a computational workload of tens or even hundreds of exaFLOPs ($10^{18}$ floating point operations) [6]. In the following we discuss the current development priorities and our strategy to meet the exascale challenges.

**Development Priorities**

(i) *Adaptation of the WEST code to heterogeneous computing*. The rise of heterogeneous computing has substantially increased the throughput available in leadership high-performance computing systems, due to the single instruction multiple threads (SIMT) parallelism introduced by modern general-purpose GPUs. Such massive and hierarchical parallelism requires a careful design of the algorithms and of the data structures used in WEST to carry out MBPT calculations. As discussed below, we are leveraging SIMT parallelization by refactoring the data and loop distribution in the code. A current challenge is to extend the performance gains, initially achieved on NVIDIA GPUs for $G_0W_0$ calculations [6], to all features of the code and using devices of other brands, as their GPU software toolchains and hardware become available. Code refactor and optimization will allow us to leverage high-performance computing systems and enable the study of complex heterogeneous materials, e.g., defects and interfaces in oxide materials; for these systems an accurate description of the single particle wavefunctions that serve as a starting point of MBPT calculations is expected to require the use of hybrid functionals, which are computationally more demanding than semi-local ones.

(ii) *Connecting different levels of theory using quantum embedding*. Numerous interesting problems in materials science and chemistry require the description of a small portion of the entire system at a level of theory higher than the rest of the material. For example, this may be the case for defects in solids, molecules undergoing chemical reactions at surfaces or nanoparticles embedded in matrices. There are multiple reasons for the need of a higher level of theory for a specific region, one being, for example, the presence of highly correlated electronic states localized in space and energy in a solid, that cannot be described using mean-field theories such as DFT. The development of embedding theories to describe different portions of a complex system at different levels of theory is an active field of research. We have recently developed the quantum defect embedding theory (QDET) based on Green's functions methods [4], where an active space is defined and an effective Hamiltonian is diagonalized exactly to obtain correlated many-body states. The extension of this method to systems other than covalently bonded semiconductors, e.g., oxides and aqueous interfaces, requires the developments of methods to include vertex corrections and achieve self-consistency in the Green's function description of the entire system, or the application of hybridization schemes between the active space and the environment.

(iii) *Code interoperability*. Workflows used in electronic structure calculations are increasing in complexity, as theoretical methods advance and the computational power grows. It is therefore of great importance to develop flexible and extensible workflows where one or more codes cooperate through in-vivo or ex-vivo coupling [7], possibly operating on different architectures, including both classical (CPUs and GPUs) [6] and quantum computers (QPUs) [8]. We have recently developed several coupling schemes to improve the efficiency of MBPT-based calculations and facilitate the development of coupled codes, as shown in Fig. 1. For example, we developed a coupling scheme between WEST and the first-principles molecular dynamics code Qbox to improve the efficiency of the solution of the BSE by performing electronic structure calculations in finite electric field, while taking advantage of electronic orbital localization. In addition, QDET calculations are performed by coupling a DFT engine (Quantum ESPRESSO or Qbox), a MBPT solver (WEST), and diagonalization codes for the effective Hamiltonian. The latter are based on quantum chemistry methods and may be run on classical or quantum architectures. As the next generation of supercomputers will be more modular and heterogenous, and the size and fidelity of quantum computers continue to improve, code coupling represents an interesting avenue to combine the strengths of diverse types of computing paradigms [8].

(iv) *Introducing machine learning protocols in electronic structure theory*. Advances in machine learning and deep learning techniques have substantially improved the efficiency of several electronic structure methods. We recently applied data-driven approaches to the calculation of the dielectric screening, a key ingredient required to compute absorption spectra using the BSE [9]. In specific cases (water and some aqueous interfaces), we obtained a model for the screening that outperforms the direct calculation of dielectric matrices by one to two orders of magnitude, while retaining transferability across multiple configurations extracted from first-principles molecular dynamics simulations. Some of the challenges in improving this technique include the updating procedure of the screening necessary in molecular dynamics simulations and its extension to compute properties other than absorption spectra. Overall, we expect that machine learning may help reduce the cost of first-principles methods by identifying redundant calculations or by providing surrogate models of complex quantities, e.g., response functions. Hence, defining protocols to rigorously verify and validate machine learning models is as important as developing the models themselves.

**Meeting the Exascale Challenges**

High performance computing (HPC) has recently entered in earnest the exascale era, with new opportunities and challenges for the electronic structure community. The throughput of HPC systems keeps increasing at a rapid pace, promising the feasibility of atomistic and first-principles simulations at unprecedented scales. However, HPC systems are becoming heterogeneous, with most of the performance currently contributed by GPU accelerators. Fully harnessing the parallelism available in leadership HPC systems, and specifically SIMT parallelism, mandates a redesign of most of the code architecture and even of the underlying algorithms. A cornerstone algorithm in WEST is the projective dielectric eigenpotentials (PDEP) method that determines the spectral decomposition of the static dielectric matrix at zero frequency by an iterative diagonalization [1]. This algorithm lends itself to efficient parallelization. The computation of the density response to multiple perturbations can be carried out fully in parallel for each perturbation, spin channel, and wavefunction, making the PDEP method scalable to over ten thousand CPU cores [1] and GPUs [6].

Code optimization is another key step required to maximize performance on heterogeneous HPC architectures. Initially written for compute nodes based on multi-core CPUs, WEST has been extended to use accelerators, starting from NVIDIA GPUs [6]. A significant speedup over the CPU version of WEST was achieved by utilizing high-performance GPU libraries, overlapping computations with communications, and mixed-precision (FP32/FP64) as GPUs are particularly efficient with reduced precision throughput. The GPU version of WEST was demonstrated to scale to 25,920 GPUs of the Summit supercomputer, reaching a mixed-precision performance of 58.8 PFLOP/s for full-frequency $G_0W_0$ calculations (see Fig. 2 and Fig. 3). Work is under way to achieve performance portability targeting the exascale supercomputers powered by NVIDIA, AMD, and Intel GPUs. To achieve this goal, we are following the principle of "separation of concerns", that is, making the code modular and relying as much as possible on existing infrastructure of libraries optimized and maintained by domain experts. In addition, we take advantage of containerization techniques to facilitate the testing of the code and its deployment on leadership HPC systems as well as mid-size clusters and workstations. A challenging activity is to deliver code that is optimized for multiple architectures while keeping the structure of the code simple and extensible, so as to avoid barriers for future developments and facilitate contributions from scientists not familiar with cutting-edge coding paradigms.

We expect the coupling of different codes running on different architectures, including quantum computers and the use of machine learning methods will play an important role in exascale computing. The GPU accelerators, which will power the announced exascale architectures, are optimized for machine learning workloads. Last but not least, we mention the importance of data collection, analysis, and curation in the exascale computing era. The unprecedented computational power will generate an unprecedented amount of data. Handling these data may be more challenging than generating them. In recent years, community efforts from all over the world have started to build data infrastructures serving the electronic structure community. Our contribution to these efforts is Qresp [10], a tool to facilitate reproducibility in science by curating data associated with scientific publications.

**Concluding Remarks**

We have presented the state-of-the-art implementation of calculations based on many-body perturbation theory using the WEST code, which comprise calculations of electron-electron, electron-hole and electron-phonon interactions of large heterogeneous systems, as well as of strongly correlated active regions of condensed systems. Development priorities include adapting the WEST code to heterogeneous computing to enable the modeling of increasingly complex heterogenous systems; refining embedding techniques to broaden their applicability to wider classes of systems; increasing the coupling of WEST with other codes and libraries to enable the use of modular workflows on exascale and quantum computers; and using methods based on machine learning to speed up electronic structure calculations.

Progress towards exascale computing was achieved by focusing on the porting of the full-frequency $G_0W_0$ solver on NVIDIA GPUs. Strong and weak scaling was demonstrated for a system that contains ~10k electrons on up to 25,920 GPUs of the Summit supercomputer, leading to a mixed FP32/FP64 performance of 58.8 PFLOP/s. The porting to NVIDIA GPUs led to a refactored code, that can be used as base line to port the performance of the code to GPUs of other vendors, as they become available.

We conclude by noting that while exascale computing is pushing the envelope of both theory and simulation, a vast range of scientists operates on tera- and petascale resources and relies on such resources to carry out their research activity. Algorithms implemented in community open-source codes like WEST need to reconcile code maintainability with the complexity required to adapt to a variety of diverse computing resources and sweeping changes in coding paradigms. As part of code maintenance, verification and testing, including testing the code at scale, are important activities. Workforce development is essential in order to sustain growth. In particular, the training and deployment of domain scientists with a diverse combination of coding skills and proficiency in artificial intelligence and quantum computing are necessary to unlock the power of emerging and future computing.

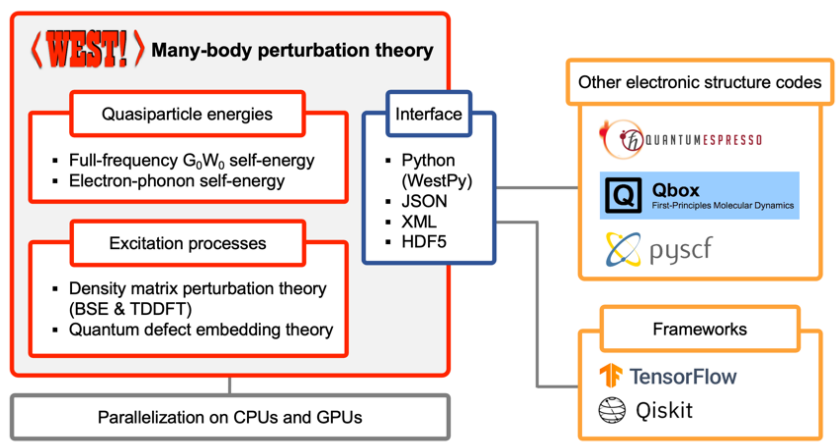

**Figure 1.** Overview of the WEST software package. Key functionalities of WEST include the computation of quasiparticle energies using full-frequency $G_0W_0$ and electron-phonon self-energies, and the description of excitation processes using density matrix perturbation theory (Bethe-Salpeter equation and time-dependent density-functional theory) and the quantum defect embedding theory. Several computational kernels of WEST have been optimized for massively parallel computers based on CPUs and GPUs. WEST utilizes a Python interface layer, WestPy, and standard data formats such as JSON, XML, and HDF5, to interoperate with other electronic structure codes [7], currently including Quantum ESPRESSO, Qbox, and pyscf. In addition, WEST can use the TensorFlow and Qiskit open-source frameworks to implement machine learning and quantum computing protocols, respectively.

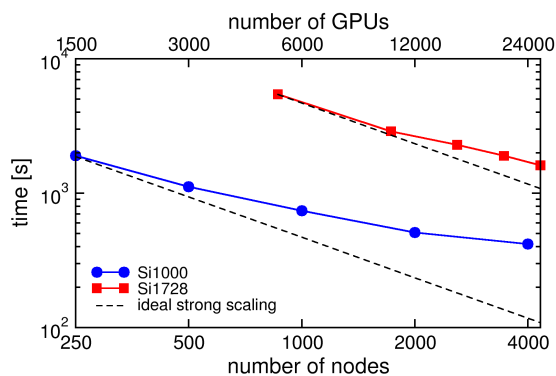

**Figure 2.** Strong scaling of the GPU version of the WEST code on the Summit supercomputer for two silicon supercells containing 1,000 atoms (blue circles) and 1,728 atoms (red squares), respectively. The black dashed lines indicate the slope of ideal scaling. Eighty quasiparticle energies (around the Fermi level, 40 below and 40 above) were calculated for each system. Timing results correspond to the total wall clock time, including the time spent on I/O operations and CPU-GPU communications. Adapted with permission from Ref. [6].

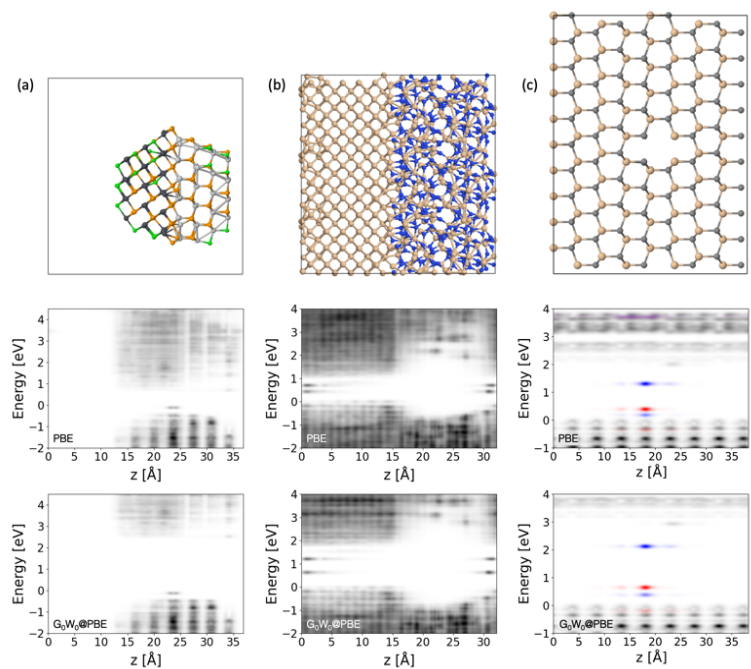

**Figure 3.** Large-scale full-frequency $G_0W_0$ calculations using the GPU version of the WEST code: (a) a Janus-like heterostructure formed by a chlorine-terminated nanoparticle made of cadmium sulfide and lead sulfide (2,816 electrons), (b) an interface of silicon and silicon nitride (10,368 electrons), and (c) a neutral hh divacancy in 4H silicon carbide (6,392 electrons, spin-polarized). The top panels show a ball-and-stick representation (side view) of the simulation cells, where Cl, Cd, S, Pb, Si, N, C atoms are represented as green, black, orange, light gray, beige, blue, and dark gray spheres, respectively. The bottom and middle panels show the local density of states (LDOS) obtained using $G_0W_0$@PBE and DFT energies, respectively. A color scale that ranges from white to black is used to represent the LDOS values; white areas indicate energy gaps. For the hh divacancy in SiC, the defect states in the up (down) spin channel are shown in red (blue). Adapted with permission from Ref. [6].

## Acknowledgements

The authors would like to thank the whole WEST development team. The authors acknowledge support provided by the Midwest Integrated Center for Computational Materials (MICCoM), as part of the Computational Materials Sciences Program funded by the U.S. Department of Energy, Office of Science, Basic Energy Sciences, Materials Sciences, and Engineering Division through Argonne National Laboratory (ANL).